\documentclass{aa}         
\usepackage{graphicx,float}
\begin{document}
\thesaurus{11     
              (09.13.2;  
               11.04.2;  
               11.07.1;  
      11.09.1  NGC\,6946;
               13.19.1;  
               13.20.1)} 
\title{Molecular gas in  NGC\,6946}
\author{W. Walsh\inst{1,4}, R. Beck\inst{1}, G. Thuma\inst{1}, A. Weiss\inst{2}, 
R. Wielebinski\inst{1} \& M. Dumke\inst{1,3}}
\offprints{W. Walsh, wwalsh@mpifr-bonn.mpg.de}
\institute{Max-Planck-Institut f\"ur Radioastronomie, 
          Auf dem H\"ugel 69, D-53121, Bonn, Germany 
\and
          Radioastronomisches Institut der Universit\"at Bonn,
          Auf dem H\"ugel 71, D-53121, Bonn, Germany 
\and
          SMTO, Steward Observatory, The University of Arizona,
          933 N.~Cherry Avenue, Tucson, Arizona 85721, USA 
\and
          Harvard Smithsonian Center for Astrophysics
          60 Garden St, MS12, Cambridge, MA 02138, USA         
          }
\date{Received ... ; accepted ...}
\authorrunning{Walsh et al.}
\titlerunning{Molecular gas in NGC\,6946}
\maketitle
\begin{abstract}

We present imaging of molecular gas emission in the star-forming
spiral galaxy NGC\,6946. Our $^{12}$CO(1-0) and $^{12}$CO(3-2) images,
made at 22\arcsec\, resolution with the IRAM 30-m and the Heinrich
Hertz 10-m radio telescopes, are the most extensive observations of
molecular gas in this galaxy and are among the most extensive
observations of molecular gas in any spiral galaxy.

The $^{12}$CO(1-0) emission shows a central concentration aligned in a
north-south ellipse, an extended diffuse component plus concentrations
in the outer spiral arms. $^{12}$CO(1-0) is detected across most of
the optical disk out to a projected radius of nearly 300\arcsec. The
molecular component in NGC\,6946 is unusually massive, with
$\mathrm{M}_{\mathrm{H}_2}/\mathrm{M}_{\mathrm{HI}} = 0.57$. The scale
length of the disk in $^{12}$CO(1-0) is the same, to within 5\%, of
the VRI, H$\alpha$, 15\,$\mu$m, and FIR disks.  The $^{12}$CO(3-2)
data shows broadly similar structure to the $^{12}$CO(1-0) image out
to the observed limit of 200\arcsec, although the arm-interarm
contrast of $^{12}$CO(3-2) is significantly larger than that of
$^{12}$CO(1-0) which suggests that molecules in the spiral arms are
warmer or reside longer there.  The rotation curve derived from the
$^{12}$CO(1-0) velocity field agrees, within the uncertainties, with
the {H\,{\sc i}}\ rotation curve.  The velocity dispersion is
$40\pm10$\,km~s$^{-1}$ in the inner 2\,kpc and $8\pm3$\,km~s$^{-1}$ in
the disk, without tendency to increase in the spiral arms.

Spectra of the $^{13}$CO(1-0), $^{13}$CO(2-1), $^{12}$C$^{18}$O(1-0)
isotopomers for several positions are used for line ratio studies of
the state of the molecular ISM in NGC\,6946.  In the centre of
NGC\,6946 our LVG analysis suggests that the beam-averaged gas kinetic
temperature is $40\pm5$~K, the molecular gas density is $(3.3\pm0.3)
\times 10^3$\,cm$^{-3}$ and that line opacities are high, with
$\tau _{^{12}\mathrm{CO}(3-2)} \sim 10$. A star formation efficiency
image for NGC\,6946, made from the H$\alpha$ image divided by the
molecular gas image, ranges by over two orders of magnitude with
highest values found in the northeastern spiral arm. The
$\lambda$6\,cm polarized emission image, which traces the regular part
of the magnetic field, appears anti-correlated with the star formation
efficiency.

We present an analysis of the ISM in NGC\,6946's disk by making 1-D
and 2-D comparisons of images made in several wavebands. Using a
point-by-point correlation technique, we investigate the distribution
and kinematics of the molecular gas and its relation to the neutral
and ionized gas, the mid-infrared-emitting dust, the radio continuum
and the magnetic field, and find that the molecular gas is closely
associated with the 7\,$\mu$m-emitting dust.  The highest correlation
between any pair of tracers is found between the mid-infrared emission
and the total radio continuum emission at $\lambda$6\,cm. This cannot
be due to dust heating and gas ionization in star-forming regions
because the thermal radio emission is {\it less} correlated with the
mid-infrared emission than the nonthermal emission.  A coupling of
magnetic fields to gas clouds is proposed as a possible scenario.

\end{abstract}
\section{Introduction}

How gas is compressed into molecular clouds of various sizes and
subsequently forms stars is one of the outstanding questions of modern
astronomy. On the scale of a galaxy, the effects of the galaxy's gas
content, chemistry, dynamical and morphological properties, spiral
structure, and magnetic field strength are among the factors that need
to be taken into account (Evans 1999). NGC\,6946 is ideally suited for
studies of these properties, as it is a nearby, nearly face--on spiral
galaxy with an exceptionally gas--rich disk which shows evidence for a
high star formation rate (SFR) throughout (de Gioia-Eastwood et
al. 1984). NGC\,6946's orientation and large angular size ($D_{25,B}
=11\farcm2$) offers the prospect of studying the structure and star
forming properties of the disk with sufficient (sub-kpc) spatial
resolution to clearly resolve the spiral arms.

NGC\,6946 and its many supernova remnants have been the subject of
numerous X--ray, optical and radio papers. Several authors have
previously studied NGC\,6946 in optical wavebands, although these
studies are hampered by the low Galactic longitude ($b = 11\fdg67$)
which results in considerable extinction ($A_B = 1.6$) and many
foreground stars.  At mm wavelengths the CO molecule, used as a
tracer of the dominant molecular species, H$_2$, has been observed by
several authors. Ball et al. (1985), Sofue et al. (1988), Weliachew et
al. (1988), Ishizuki et al. (1990) and Regan \& Vogel (1995) observed
just the centre region, while Casoli et al. (1990) imaged two
110\arcsec\, diameter regions with the IRAM 30-m telescope. Morris
\& Lo (1978), Rickard \& Palmer (1981), Young \& Scoville (1982) and
Tacconi \& Young (1989) obtained low resolution CO images of parts of
the inner disk; Sauty et al. (1995) present the largest previous
molecular image, in the CO(2-1) line.

Morphologically, NGC\,6946 is a late-type SAB(rs)cd spiral galaxy,
with several spiral arms and star-forming regions scattered throughout
the disk, considerable extinction caused by dust lanes (Sauty et
al. 1995; Trewhella 1998) and a small nucleus. Its general properties
are summarized in Table~\ref{t:general}. There is some debate in the
literature about how many spiral arms the numerous fragments of arms
belong to, but $K$-band images reveal four prominent, not very
symmetric arms (Regan \& Vogel 1995). The ``thick'' optical arm in the
NE, also seen in the deep H$\alpha$ map presented by Ferguson et
al. (1998), earns NGC\,6946 a place in Arp's atlas.

Although NGC\,6946 has only a small bar (Elmegreen et al. 1998),
Ishizuki et al. (1990) present evidence for an efficient inflow of gas
into the central 11\arcsec, fueling the activity observed there
(Turner \& Ho 1983; DeGioia-Eastwood 1985; Roy \& Belley 1993;
Elmegreen et al. 1998) which has been interpreted as a mild or post
starburst (Tacconi \& Young 1990; Engelbracht et al. 1996; Ptak et
al. 1999). Devereux \& Young (1993) discuss the origin of the bright
far-IR emission in NGC\,6946 (Telesco \& Harper 1980; Engargiola 1991)
which, at $5 \times 10^{10}\,L_{\odot}$, is amongst the highest
measured in a nearby spiral galaxy (Devereux \& Young 1993). Devereux
\& Young (1993) conclude that most of the FIR luminosity derives from
warm ($\sim 33$~K) dust although 90\% of the mass of the dust is in a
cooler ($\leq 17$~K) component.
\begin{table}
\caption[]{NGC\,6946 general properties}
\label{t:general}
\[
\begin{tabular*}{\linewidth}[htb]{lr}
\hline
\noalign{\smallskip}
Property & Value \\
\noalign{\smallskip}
\hline
\noalign{\smallskip}
Type	        	& SAB(rs)cd$^{\mathrm{a}}$                  \\
R.A. (J2000)		& 20 34 52.3$^{\mathrm{a}}$                 \\
Dec. (J2000)		& 60 09 14.2$^{\mathrm{a}}$                 \\
Distance                & 5.5 Mpc$^{\mathrm{b}}$                    \\
$D_{25}\times d_{25}$	& 11\farcm2 (16.6\,kpc) $\times 10.0$ (14.8\,kpc)$^{\mathrm{a}}$ \\
V$_{sys}$               &  $47 \pm 4$\,km~s$^{-1}$$^{\mathrm{b}}$     \\
m$_0^T$(B)              &  7.78$^{\mathrm{a}}$                      \\
M$_0^T$(B)              &  -20.92                                   \\ 
L$_0^T$(B)              &  5.3$\times 10^{10}$~L$_{B\odot}$         \\
$B-V$                   &  0.4$^{\mathrm{a}}$                       \\
M/L (B) disk            &  $1.2 \pm 0.2^{\mathrm{c}}$               \\
M$_{*}$	         &$6.4 \times 10^{10}~\mathrm{M}_{\odot} ^{\mathrm{c}}$\\
M$_{HI}$                &  $2.0\pm0.1 \times 10^{10}~\mbox{M}_{\odot}$$^{\mathrm{c,e}}$\\
M$_{H_2}$               &  $1.2\pm0.1 \times 10^{10}~\mbox{M}_{\odot}$$^{\mathrm{f}}$\\
M$_{dust}$              &  $1.6\pm0.5 \times 10^{8}~\mbox{M}_{\odot}$$^{\mathrm{i}}$\\
M$_{dm}$		&  $1.6 \times 10^{11}~\mbox{M}_{\odot}$$^{\mathrm{c,d}}$\\
At R$_{\mathrm{HO}}$ M$_{dark}$/M$_{lum}$ & $0.73^{\mathrm{c}}$ \\
Inclination             &  $38 \pm 5^{\mathrm{c}}$                   \\
Position angle          &  $240 \pm 3^{\mathrm{c}}$         	     \\
Metallicity		&  12 + log O/H = 9.36                       \\
Metal. Gradient         &  -0.089$^{\mathrm{j}}$ dex kpc$^{-1}$      \\
SFR	         	& 0.013 M$_{\odot}$\,yr$^{-1}$\,kpc$^{-2}$   \\
IRAS 12, 25, 60, 100~$\mu$m &  15, 23, 168, 363 Jy$^{\mathrm{h}}$    \\
200~$\mu$m              & 743 $\pm 223$ Jy$^{\mathrm{i}}$            \\
\noalign{\smallskip}
\hline
\end{tabular*}
\]
\begin{list}{}{}
\item[$^{\mathrm{a}}$] RC3, m$_0^T$(B) is the apparent magnitude in $B$, corrected for foreground and intrinsic extinction.      
\item[$^{\mathrm{b}}$] Tully (1988). Note that the distance is uncertain, and other authors have quoted values between 3.2 and 11~Mpc.          
\item[$^{\mathrm{c}}$] Carignan et al. (1990): R$_{\mathrm{HO}}$ = 11.6\,kpc for our adopted distance.
\item[$^{\mathrm{d}}$] Mass of the dark matter halo, measured at the last point on rotation curve (14.7\,kpc).   
\item[$^{\mathrm{e}}$] Rogstad et al. (1973)
\item[$^{\mathrm{f}}$] Tacconi \& Young (1986), Sect.~\ref{s:co.mass}.
\item[$^{\mathrm{g}}$] Burstein \& Heiles (1982): A$_B$ = 1.610 mag
\item[$^{\mathrm{h}}$] IRAS: Rice et al. (1988).  
\item[$^{\mathrm{i}}$] Alton et al. (1998). Note that this dust mass, derived from a combination of ISO 100 and 200~$\mu$m data, is an order of magnitude larger than that derived from IRAS data alone, and assumes an index of 1.0 in the FIR emissivity law.
\item[$^{\mathrm{j}}$] This value can be compared with a mean value of -0.048 for a sample of 28 spiral galaxies reported in Dutil \& Roy (1999).
\end{list}
\end{table}

Alton et al. (1998) and Bianchi et al. (2000) present images of NGC\,6946 in
the 450~$\mu$m and 850~$\mu$m wavebands, respectively, and find that
dust is associated with the molecular ISM, becoming more dominant at
larger radii, where it is possibly more extended than the stars. Dale
et al. (1999; 2000) present ISOCAM maps of NGC\,6946 in the LW2 and
LW3 bands centered at 7\,$\mu$m and 15\,$\mu$m.

NGC\,6946 has also been observed intensively in radio continuum,
including the linearly polarized emission which indicates regular
magnetic fields (Beck 1991; Beck \& Hoernes 1996). The polarized
emission emerges from two ``magnetic arms'' located between the
optical arms like their phase-shifted images (Frick et al. 2000).

Previous CO observations have identified NGC\,6946 as having one of
the most massive and extended molecular gas components observed in a
nearby galaxy (e.g. Young et al. 1995). Atomic gas is also found
throughout the disk, although it is found in highest concentrations in
the spiral arms and appears less massive than the molecular component
(Tacconi \& Young 1986; Boulanger \& Viallefond 1992). At present,
theoretical models of large scale star formation consider the
equilibrium reached between the stellar radiation field and physical
state of the gas clouds to be crucial (Parravano 1989; Bertoldi \&
McKee 1996), and thus require quantitative input on the relative
distribution of the stars and the gas. A uniform image of the entire
disk in the primary molecular gas tracer, the $^{12}$CO(1-0) line, is
therefore long overdue for NGC\,6946. Since stars form primarily in
the dense cores of molecular clouds, observing the emission from
molecules that require high densities for excitation is important for
estimating the density, mass, temperature and opacity of star-forming
clouds in galaxies. We therefore also present extensive imaging of the
$^{12}$CO(3-2) line, which may be only moderately opaque (Petitpas \&
Wilson 2000), and therefore trace the column density of the warm
component better than other lines.

The distance to NGC\,6946 has been variously quoted to lie within the
range 3.2-11~Mpc. We adopt a value of 5.5~Mpc (Tully 1988), which
means that our $\sim$22\arcsec\, resolution corresponds to a linear
scale of 0.59\,kpc.

Sect.~\ref{s:observations} describes the IRAM and HHT observations, the
results of which are presented in Sect.~\ref{s:results}, which also
includes a large velocity gradient analysis of the ISM at selected
positions. Sect.~\ref{s:largescale} considers both 1-D and 2-D
correlations between images of NGC\,6946 made in several wavebands,
and Sect.~\ref{s:conclusion} summarizes the main results.

\section{Observations\label{s:observations}}

\begin{table*}[htb]
\caption[]{Receiver characteristics}
\label{t:receivers}
\[
\begin{tabular*}{\linewidth}[h]{lccccc}
\hline
\noalign{\smallskip}
Receiver & Polarizations & Tuning range & Receiver temperature & Beam efficiency & Forward efficiency  \\
 &  &  (GHz)  &  (K) & B$_{\mathrm{eff}}$ & F$_{\mathrm{eff}}$ \\
\noalign{\smallskip}
\hline
\noalign{\smallskip}
SORAL 230~GHz SIS & 1 & 210-275 & 60-80$^{\mathrm{a}}$ & 0.77 & 0.9\\
MPIfR 345~GHz SIS & 2$^{\mathrm{b}}$ &  320-375 & 80--140$^{\mathrm{a}}$ & 0.45-0.5 & 0.9\\
SORAL 490~GHz SIS & 1 & 435-498 & 130--180$^{\mathrm{a}}$ & 0.45 & 0.72\\
\noalign{\smallskip}
\hline
\end{tabular*}
\]
\begin{list}{}{}
\item[$^{\mathrm{a}}$] Double sideband
\item[$^{\mathrm{b}}$] Beam misalignment $\leq$ 5\arcsec
\end{list}
\end{table*}

\subsection{Pico Veleta observations}

The IRAM 30-m telescope, located at elevation 2920~m on Pico Veleta,
Spain, is a paraboloid with f/D=0.35.  The telescope uses Nasmyth foci
and the main dish surface is accurate to $70\pm5~\mu$m rms. We
observed in ``on-the-fly'' mode where the telescope beam moves with a
uniform velocity across the source. Orthogonal raster maps were made
of a 10\arcmin~$\times$~10\arcmin\, field. The integration time per
point (on source) was 4\,s and the maps were repeated three times in
each direction and then averaged. One 1024-channel filterbank was used
to observe CO(1-0) with a resolution of 1~MHz or 2.6~km~s$^{-1}$ and
an autocorrelator was used for CO(2-1). Both frequencies were
observed simultaneously and data were recorded every 1s. Typical
system temperatures at 115\,GHz were $260\pm20$~K. The beamwidths of
the 30-m telescope at 115 and 230~GHz are 21\arcsec\, and 12\arcsec,
respectively.  To (more than) satisfy the sampling theorem for the
CO(1-0) map we used 9\arcsec\, raster spacing. An emission-free
reference position was measured every horizontal or vertical raster,
or about once every 7.5 minutes.

\begin{figure}[bth]
\resizebox{\hsize}{!}{\includegraphics*[angle=-90,totalheight=6cm]{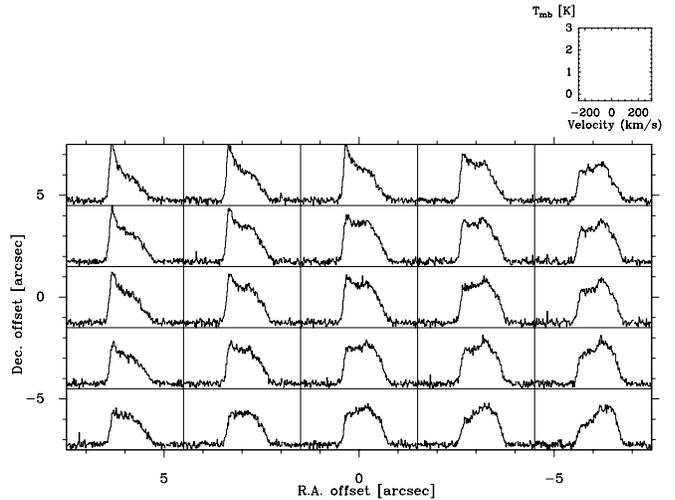}}
\caption[]{$^{12}$CO(1-0) spectra of the central positions in NGC\,6946, 
made with the IRAM 30-m radio telescope, which has a FWHM at 110~GHz 
of 22\arcsec. Pointing centres are spaced by 3\arcsec. 
 \protect{\label{f:pointing}}}
\end{figure}

Calibration, performed with the chopper wheel method with both hot
(sky and absorber) and cold (nitrogen) loads, is described in detail
in Wild (1999). The conversion factor from main beam brightness
temperature (T$_{\mathrm{mb}}$) to flux density is
4.7\,Jy~K$^{-1}$. The calibration was verified on well-known molecular
sources and by continuum scans across Jupiter, Uranus, Mars and
Saturn, and found to be better than 10\% at both frequencies. At
230~GHz problems with the autocorrelator compromised data quality and
these data are not discussed further. The absolute pointing model is
checked and modeled regularly by the IRAM staff so that our focus and
pointing checks, made every few hours and after sunrise/sunset or long
drives, required corrections of always less than 4\arcsec. Tracking
precision of the 30-m telescope is better than 1\arcsec.

Fig.~\ref{f:pointing} shows $^{12}$CO(1-0) spectra of the central
positions in NGC\,6946 with pointing centres spaced by 3\arcsec. Since
these central spectra in NGC\,6946 are so bright and distinctive, they
were frequently used as an additional check on the pointing for both
the IRAM and the HHT observations.

The data were reduced with the CLASS and GRAPHIC software of the
Grenoble Astrophysical Group (GAG) package. A polynomial baseline of
order one, or occasionally three, was removed from each spectrum and
then spectra taken towards the same position were averaged. As several
coverages of the galaxy (in orthogonal directions) had been taken, the
different maps were compared before averaging, to ensure consistency
between them.

Further processing of the on-the-fly data to remove the some scanning
artifacts was done using the algorithm described by Emerson \& Gr\"ave
(1988). We used the FLYPLAIT implementation of this algorithm as
described by Hoernes (1997) and Nieten (2001), which computes the FFT
of the two sets of orthogonal scans and then performs a weighted
addition in the Fourier transformed domain. The inverse FFT images
have slightly lower noise levels and are free of residual scanning
image defects. The resulting images were compared with comparably
regridded and smoothed images to ensure that the FFT process did not
introduce new features, flux offsets and the like. Finally, channel
maps and integrated maps in suitable velocity intervals were created.
  
\subsection{Heinrich Hertz Telescope observations}

The CO(3-2) observations were performed with the Heinrich Hertz
Telescope \footnote[1]{The HHT is operated by the Submillimeter
Telescope Observatory on behalf of Steward Observatory and the MPI
f\"ur Radioastronomie.}  (Baars et al. 1999), located at elevation
3178~m on Mt Graham, Arizona, during 1999 April, 1999 November, 2000
January and 2000 November. The HHT is a 10~m paraboloid with f/D=0.35,
Nasmyth foci and surface accuracy better than $12\pm3~\mu$m rms
(i.e. $\lambda/66$ at 345~GHz). The FWHM beam sizes are
$16\farcs5\pm1$\arcsec\,, $22\arcsec\pm1$\arcsec\, and
$36\arcsec\pm1$\arcsec\, at 490~GHz, 345~GHz and 230~GHz,
respectively, while the absolute pointing accuracy is 1--3\arcsec\,
rms and the tracking precision is 0\farcs2 rms (SMTO Electronic
Newsletter 1998).

The backends used were two acousto-optical spectrometers, each with a
total bandwidth of 1~GHz. For the CO(3-2) line observations one AOS
was used for each linear polarization channel. The 345~GHz receiver is
sensitive to both sidebands and the CO line was tuned to the lower
sideband. The secondary mirror was `wobbled' with a beam throw of
240\arcsec\, in azimuth and scans obtained with emission--free
reference positions on either side were coadded to ensure flat
baselines.

All observations in the CO(3-2) line were performed with $\tau_{\rm
225~GHz} \leq 0.2$; poorer conditions were used for CO(2-1)
measurements. CO(4-3) observations were made when $\tau_{\rm 225~GHz}
\leq 0.06$. The atmospheric fluctuations and the atmospheric model
introduce a calibration uncertainty not greater than 5\%. The HHT has
no strong error beams at 345~GHz. This reduces the effect of source
size on the appropriate source coupling efficiency. The forward
efficiency, $F_{\mathrm{eff}}$, is close to 1, and the uncertainty in
$B_{\mathrm{eff}}$ introduces a 6\% calibration uncertainty.

Pointing is controlled in software using a model determined by the
observatory staff, and is based on measurements of temperature,
elevation and the sky opacity at 230~GHz, which is obtained from a
tilting radiometer. We made pointing observations of strong continuum
sources at regular intervals and after sunset/sunrise, shutdowns and
long telescope drives. Corrections obtained from these measurements
suggest that our pointing was better than 4\arcsec\, in 1999 November
and 2000 January. For a 22\arcsec\, beam, the resulting calibration
uncertainty could result in an underestimate of the intensity of a
point source of 9\%, and less for extended sources.

Our calibration strategy follows that described in Sect.~8.2.5 of
Rohlfs \& Wilson (1996). After retuning and at frequent other
intervals we measured the receiver temperature using an ambient load
and a load at liquid nitrogen temperature. This was used to scale the
online temperatures, which after baselines of order one (and, very
occasionally, three) were subtracted to obtain our estimate of ${\rm
T}^*_{\rm A}$. We relate these observed values to a main-beam
brightness temperature ${\rm T}_{\rm mb}$ using ${\rm T}^*_{\rm mb} =
{\rm F}_{\rm eff}/{\rm B}_{\rm eff} {\rm T}^*_{\rm A}$ using the
efficiencies listed in Table~\ref{t:receivers}.  Any imbalance of the
gains in the lower and upper sideband will increase the calibration
errors as the hot/cold calibration signals and the atmospheric and
receiver noise are present in both bands while we measure the line in
only one band. Therefore in practice we measure standard Galactic line
sources immediately before, during or after NGC\,6946 and scale ${\rm
T}^*_{\rm A}$ accordingly. For the 345~GHz receiver we do this
separately for the two channels.

Our calibration strategy requires reliable standard line strengths for
Galactic objects distributed over the sky. Unfortunately such a
catalog does not exist at all submillimetre frequencies. Nevertheless
we observed a few standard sources as often as was practicable, and
included in every significant period of observing time an observation
of the bright, distinctive central profile of NGC\,6946. Our principal
standard sources are $\Omega$~Ceti, $\chi$~Ceti and IRC~10216 (Wang et
al. 1994; Stanek et al. 1995; Young 1995; Knapp et al. 1995) and those
in Mauersberger et al. (1999).

Although the $^{12}$CO(3-2) observations are the result of several
observing sessions at widely separated epoch, the inter-epoch
calibration reliability is much higher than the absolute
calibration. Furthermore, observations were repeated many times so
that lines at all positions are detected with a significance of
$3\sigma$ or more.
\section{Results\label{s:results}}
\subsection{Morphology of the molecular gas}
\begin{figure*}[t]
\includegraphics*[angle=0,totalheight=16cm]{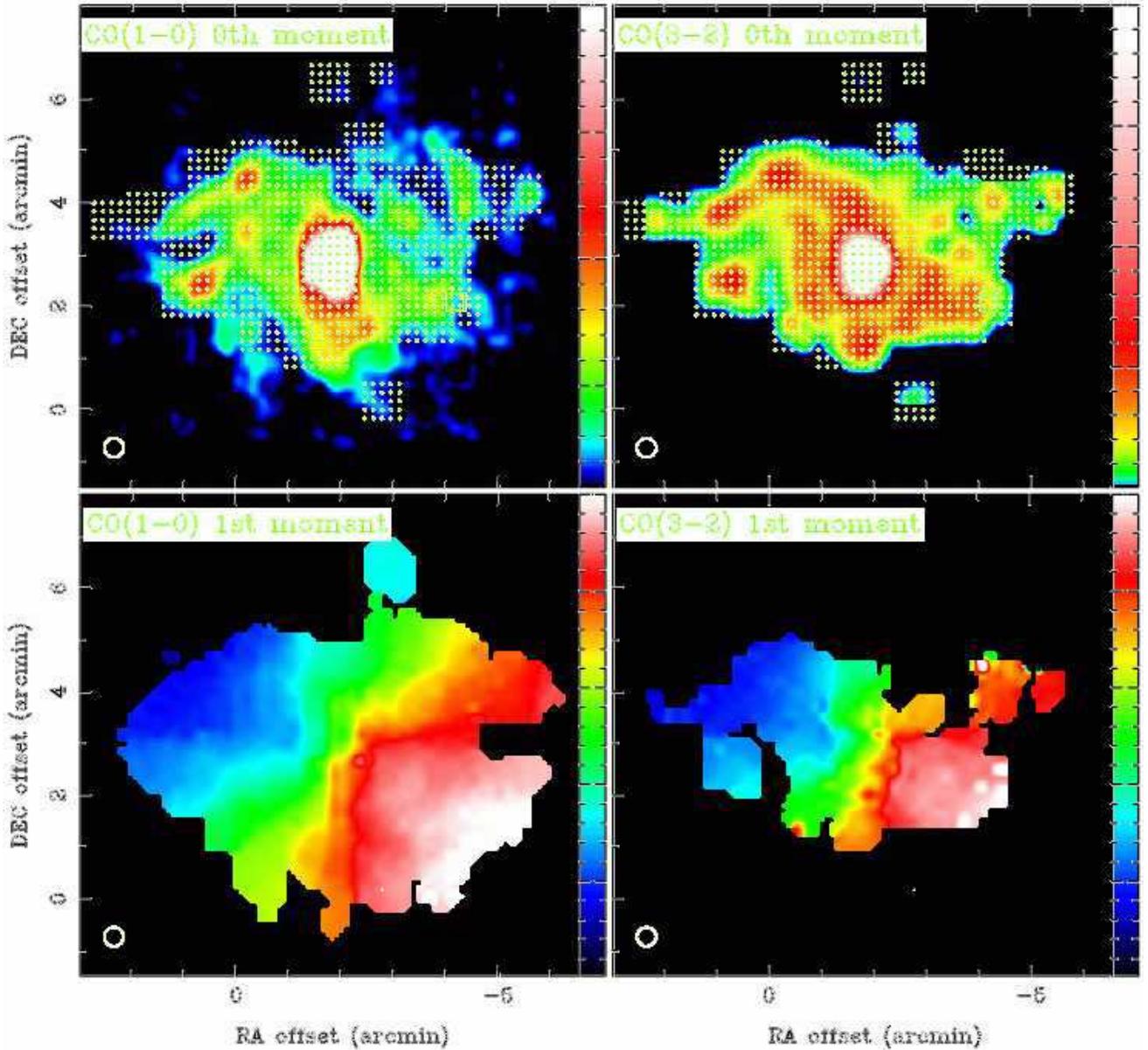}
\hfill \parbox[b]{17.5cm}
{\caption[]{\protect{\label{f:co.data}} Zeroth moment (integrated
intensity) images of the $^{12}$CO(1-0) and $^{12}$CO(3-2) spectra of
NGC\,6946 together with first moment (velocity field) images. In the
lower left hand corner of each plot the beam sizes are indicated. For
the zeroth moment $^{12}$CO(1-0) image, the colour scale ranges from
3\,$\sigma$ to 20\,$\sigma$ or 150\,mK\,km\,s$^{-1}$ to
3\,K\,km\,s$^{-1}$.  For the zeroth moment $^{12}$CO(3-2) image the
colour scale ranges from 3\,$\sigma$ to 10\,$\sigma$ or
300\,mK\,km\,s$^{-1}$ to 3\,K\,km\,s$^{-1}$. Both use the
T$_{\mathrm{mb}}$ scale. For both velocity fields, the colour scale
ranges from -45 to 150\,km\,s$^{-1}$ with higher velocities being in
the southwest. The small crosses in the integrated $^{12}$CO(1-0)
image indicate the positions where we looked for $^{12}$CO(3-2)
emission, and the box indicates the location of the young globular
cluster reported by Elmegreen et al. (2000).}}
\end{figure*}
\begin{figure*}
\includegraphics*[angle=0,totalheight=4.8cm]{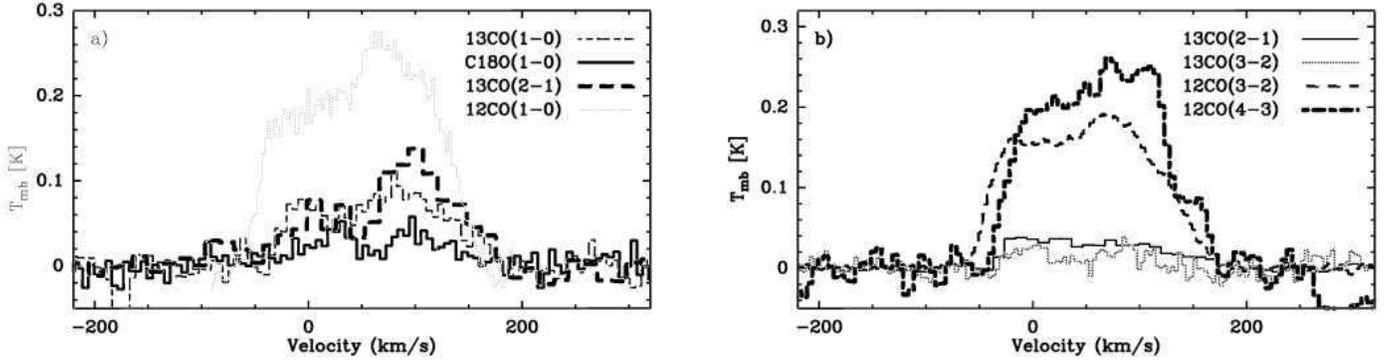}
\hfill \parbox[b]{17.5cm}{\caption[]{Spectra for the central position of 
NGC\,6946 obtained with the IRAM 30-m telescope (a) and the Heinrich
Hertz Telescope (b). In a) are shown $^{13}$CO(1-0), $^{13}$CO(2-1),
$^{12}$C$^{18}$O(1-0) and, for comparison the $^{12}$CO(1-0) line,
multiplied by 0.25.  In b) are shown $^{13}$CO(2-1), $^{13}$CO(3-2),
$^{12}$CO(4-3) and, for comparison the $^{12}$CO(3-2) line, multiplied
by 0.25. Note that the beam sizes for the CO(2-1) and CO(4-3) observations differ from the other two transitions (c.f. Table~\ref{t:spectra}).}}
\label{f:multiline}
\end{figure*}

Fig.~\protect{\ref{f:co.data}} shows the main observational results of
this study: the $^{12}$CO(1-0) and $^{12}$CO(3-2) images of NGC\,6946
and the velocity fields derived from these datasets. For the purposes
of line ratio studies, observations of selected regions were also made
in the $^{13}$CO(1-0), $^{13}$CO(2-1), $^{13}$CO(3-2), $^{12}$CO(4-3)
and $^{12}$C$^{18}$O(1-0) lines. A selection of these spectra is shown
in Fig.~\ref{f:multiline} and listed in Table~\ref{t:spectra}: other
positions will be the subject of a forthcoming paper.

\begin{table}[htb]
\caption[]{Pointed observations of various lines in NGC\,6946}
\label{t:spectra}
\[
\begin{tabular*}{\linewidth}[hbt]{lcrrrr}
\hline
\noalign{\smallskip}
Line          & Position$^{\mathrm{a}}$ & ${\rm T}_{\rm mb}$ & S/N$^{\mathrm{b}}$ & Beam$^{\mathrm{c}}$ & Corr.$^{\mathrm{d}}$\\
\noalign{\smallskip}
\hline 
$^{12}$CO(1-0)&     0    0              & 1.10               &10.0 &  21.0  &   mapped \\
$^{13}$CO(1-0)&     0    0              & 0.09               & 5.0  &    22.0   &   0.99\\
$^{13}$CO(2-1)&     0    0              & 0.036             & 7.0   &   12.6   &   mapped \\
$^{12}$CO(3-2)&     0    0              & 0.76               & 9.0  &    22.0   &   mapped \\
$^{13}$CO(3-2)&     0    0              & 0.03               & 2.8  &    23.0   &   mapped \\
$^{12}$CO(4-3)&     0    0              & 0.55               & 9.0   &   16.5   &   mapped \\
$^{12}$C$^{18}$O(1-0)&     0    0   & 0.02               & 3.0   &   22.0   &   0.99   \\
$^{12}$CO(1-0)&    20    0              & 1.10               &13.0 &    21.0  &   mapped \\
$^{13}$CO(1-0)&    20    0              & 0.09               & 4.0   &   22.0   &   0.99   \\
$^{13}$CO(2-1)&    20    0              & 0.07               & 5.0  &    12.6   &   0.93   \\
$^{12}$CO(3-2)&    20    0              & 0.40               & 9.0  &    22.0   &   mapped \\
$^{12}$CO(4-3)&    20    0              & 0.20               & 7.0  &    16.5   &   mapped \\
$^{12}$C$^{18}$O(1-0)&    20    0   & 0.01               & 3.0   &   22.0   &   0.99   \\
$^{12}$CO(1-0)&   110  100            & 0.60               & 9.0   &   21.0   &   mapped \\
$^{13}$CO(1-0)&   110  100            & 0.06               & 5.0   &   22.0   &   0.99   \\
$^{13}$CO(2-1)&   110  100            & 0.07               & 5.0   &   12.6   &   0.93   \\
$^{12}$CO(3-2)&   110  100            & 0.25               & 8.0   &   22.0   &   mapped \\
$^{12}$C$^{18}$O(1-0)&   110  100 & 0.01               & 3.0  &    22.0   &   0.99   \\
\noalign{\smallskip}
\noalign{\smallskip}
\hline
\end{tabular*}
\]
\begin{list}{}{}
\item[$^{\mathrm{a}}$] Offset in arcseconds relative to $\alpha$ = 20:34:52.3 (J2000) $\delta$ = 60:09:14.2 (J2000).
\item[$^{\mathrm{b}}$] All signal-to-noise values S/N were calculated from 
spectra with velocity resolution 10\,km\,s$^{-1}$.
\item[$^{\mathrm{c}}$] Beamsize in arcseconds.
\item[$^{\mathrm{d}}$] Beamsize correction factor for regions not mapped and smoothed.
\end{list}
\end{table}

The CO images in Fig.~\ref{f:co.data} are presented so that only
emission detected with a significance of 3$\sigma$ or more is
displayed. In both the $^{12}$CO(1-0) and $^{12}$CO(3-2) images, the
bright inner region around the nucleus and bar contains far the
brightest emission. This feature is extended north-south in both
images, approximately in line with the inner molecular bar, reported
to have a position angle of $140\degr$\, by Ishizuki et al. (1990) and
Regan \& Vogel (1995). The region of central concentration is not
symmetric, extending further southwards and westwards. As is the case
in the MIR images of Dale et al. (1999; 2000), our radio continuum
images, the [CII] images of Madden et al. (1993) and the X-ray images
of Schlegel (1994) and Schlegel et al. (2000), a diffuse emission
component is also evident in addition to concentrations in the centre
and in the vicinity of the spiral arms. The diffuse $^{12}$CO(1-0)
emission is stronger and more extended on the eastern side: a square
area of 9\,kpc$^2$ immediately east of the nucleus (excluding the NE
spiral arm) has 2.0 times as much $^{12}$CO(1-0) emission as the same
area west of the nucleus.

The $^{12}$CO(3-2) line is generally thought to trace a warmer and
denser component of the ISM than the $^{12}$CO(1-0), and thus one
might expect it to be concentrated in the spiral arms and in other
regions of current and recent active star formation (SF). Indeed the
centre of the $^{12}$CO(3-2) image contains by far the strongest
emission, and strong emission is also seen in several locations that
are also strong in H$\alpha$ and radio thermal continuum
(c.f. Fig.~\ref{f:all.data}). The two regions of most prominent
$^{12}$CO(3-2) emission are both located in the eastern spiral arms
(at $\alpha = 20^{\rm h}\, 35^{\rm m}\, 05.6^{\rm s}, \delta =
60^{\circ} 10\arcmin\, 53\arcsec\, $ and $\alpha = 20^{\rm h}\,
35^{\rm m}\, 12.5^{\rm s}, \delta = 60^{\circ} 08\arcmin\, 54\arcsec\,
$ (J2000)). Both of these ``hotspots'' are also prominent in the ISO 7
and 15\,$\mu$m images. This suggests that the corresponding emission in
the ISO images is indeed due to dust associated with gas, and not due
to K giant stars.

The $^{12}$CO(3-2) image is less extensive than the $^{12}$CO(1-0)
image since less of the disk was observed than was possible at the
lower frequency. The $^{12}$CO(3-2) images also contains considerable
diffuse emission between the hotspots and the spiral arms indicating
that, as in the Galaxy (Solomon et al. 1985), molecular clouds
containing warm and dense gas are distributed {\em throughout} the
inner disk of NGC\,6946. Detection of such extended $^{12}$CO(3-2)
emission is a novel and significant result. Until recently,
observations of higher--J CO transitions have generally been
restricted to the central regions of galaxies, where conditions of
dense gas and intense SF, which tends to heat the gas by way of
radiation from young stars, are found. Figure~\ref{f:co.data} shows,
however, that the $^{12}$CO(3-2) line is detectable over several
kiloparsecs both within the spiral arms in NGC\,6946 and throughout
the inner disk generally. In a series of recent observations
Wielebinski et al. (1999), Nieten et al. (1999), Dumke et al. (2001)
and Walsh et al. (2002) also report the discovery of emission in the
$^{12}$CO(3-2) line extended throughout the disks of several spiral
and irregular galaxies.

The strongest $^{12}$CO(1-0) emission in the outer disk coincides with
the four main spiral arms seen in $K$- and $R$-band optical
images. Two particularly bright regions occur in two of the spiral
arms east of the nucleus (around $\alpha$ = 20:35:12.3 $\delta$ =
60:08:53.1 (J2000) and $\alpha$ = 20:35:26.8 $\delta$ = 60:11:05.1
(J2000)). Both of these regions are coincident with bright H$\alpha$
emission and peaks in the $^{12}$CO(3-2) image. The $^{12}$CO(1-0)
emission is not especially strong in the vicinity of the the location
of the young globular cluster reported by Elmegreen et al. (2000), the
molecular gas having perhaps been consumed or dissociated by the
formation of the cluster.

Interarm contrast is an important quantity for constraining spiral
density wave models, but it is difficult to quantify as the definition
of where the arms end is problematic. However, NGC\,6946 is relatively
face--on and has distinct interarm regions larger than our CO
beamwidth where the optical contrast is high. In $^{12}$CO(3-2) the
arm--interarm contrast, measured over two 3\,kpc$^2$ areas located
inside and immediately south of the NE spiral arm, is $1.8 \pm 0.2$,
while over the same areas the contrast in $^{12}$CO(1-0) is $1.2 \pm
0.2$, similar to that of the {H\,{\sc i}}\ gas (Fig.~\ref{f:all.data})
which is $1.4 \pm 0.1$ over the same area. The ratio of the maximum to
minimum {H\,{\sc i}}\ column density is greater than 4.5 in the NE
arm, whereas the $^{12}$CO(1-0) emission varies only by about a factor
of 2.4 over the same regions of maximum {H\,{\sc i}}\ interarm
contrast (for example, around $\alpha$ = 20:35:12.7 $\delta$ =
60:09:36.8 (J2000)). The $^{12}$CO(3-2) maximum to minimum variation
in this region is about 5.9. If molecular hydrogen is not being
converted into or created from atomic gas (c.f. Tilanus \& Allen
1989), then continuity considerations require that the ratio of the
mass in the spiral arms to the mass in the interarm regions equals the
ratio of the time the gas spends in each. The similar contrasts in
$^{12}$CO(1-0) and {H\,{\sc i}}\ gas may indicate that the molecular
gas is forming from the atomic in these regions of the disk. The
significant difference in the $^{12}$CO(1-0)--$^{12}$CO(3-2)
arm-interarm contrasts seen in NGC\,6946 suggests that molecules are
warmer or reside for longer in the spiral arms.

\begin{figure}[htb]
\resizebox{\hsize}{!}{\includegraphics*[angle=0,totalheight=5cm]{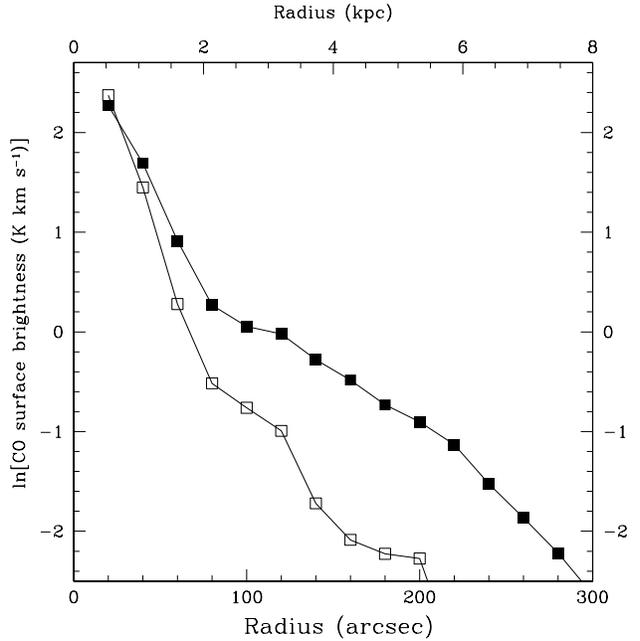}}
\caption[]{Radial distribution of molecular gas in NGC\,6946 taken from our 
$^{12}$CO(1-0) data (solid squares) and $^{12}$CO(3-2) (open squares). 
}
\label{f:co.r}
\end{figure}

The radial distribution of $^{12}$CO(1-0) in Fig.~\ref{f:co.r} was
produced by averaging in concentric elliptical annuli, using the disk
orientation parameters from the kinematical analysis of Carignan et
al. (1990). The azimuthally--averaged radial $^{12}$CO(1-0)
distribution shows how the molecular emission is strongly concentrated
towards the centre, with a steep incline within the inner 2\,kpc.

\subsection{CO kinematics} 

\begin{figure*}
\includegraphics*[angle=0,totalheight=19.2cm]{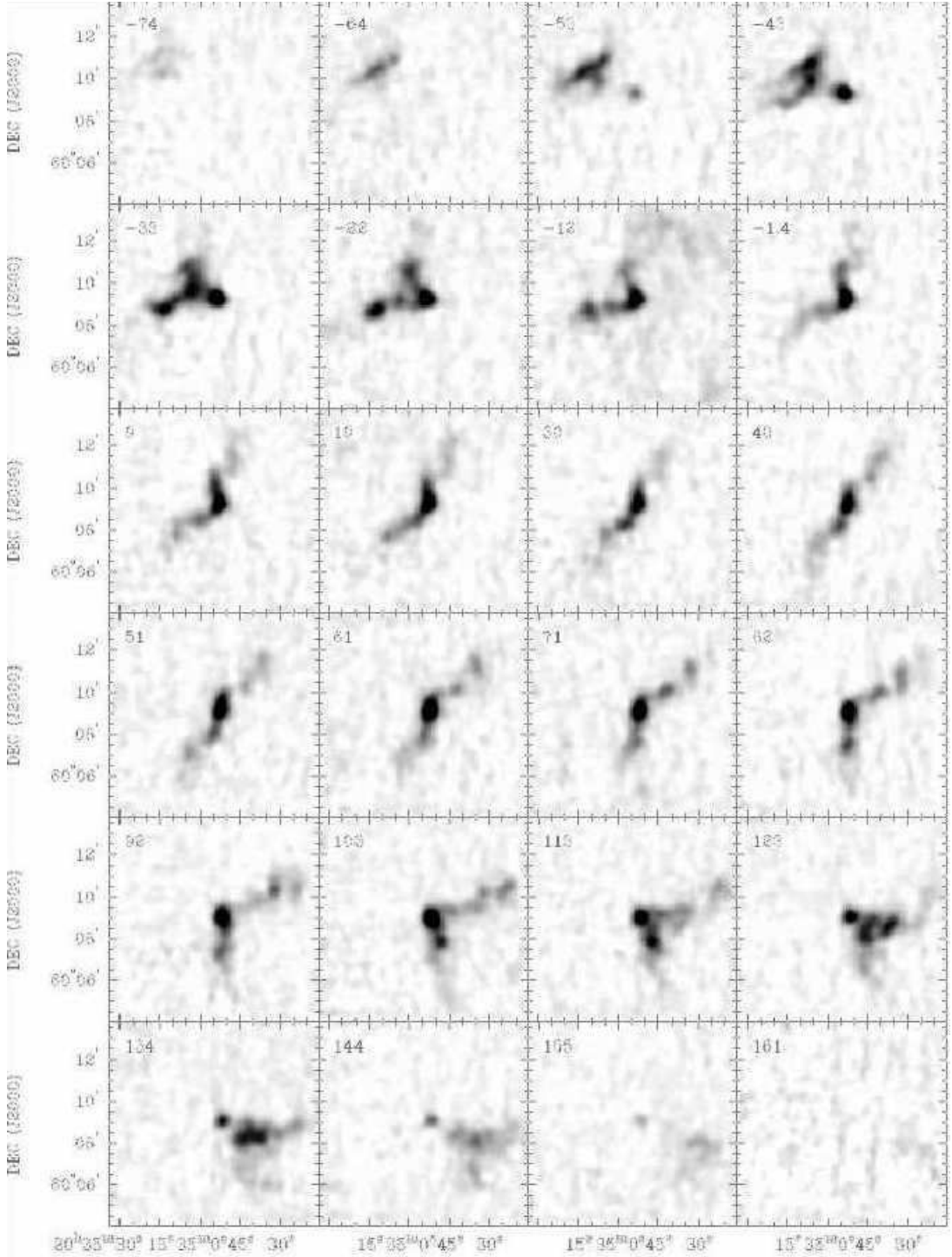}
\hfill \parbox[b]{15.5cm}{
\caption[]{Channel maps for 24 channels of the $^{12}$CO(1-0)
containing emission in the data cube made with 10\,km\,s$^{-1}$ channel
spacing. The central velocity in km\,s$^{-1}$ of each channel is
indicated in the top left hand corner of each
image.\label{f:co10chmaps}
}}

\end{figure*}

The $^{12}$CO(1-0) velocity field (Fig.~\protect{\ref{f:co.data}})
exhibits the large-scale signature of a rotating disk, as does
Fig.~\ref{f:co10chmaps}, which shows images of NGC\,6946 made in 24
channels, each approximately 10\,km\,s$^{-1}$ wide. Some departures from
circular motion are evident in the shape of the isovelocity contours,
particularly northeast of the nucleus (also the channel at
-43\,km\,s$^{-1}$). These irregularities in the velocity field may be
associated with the density wave in the northern
arm. Figure~\ref{f:co10chmaps} shows that, in the inner 590\,pc, emission
is detected over a deprojected range of 336\,km\,s$^{-1}$. A
steeply--rising central CO rotation curve, is also seen in the nuclear
disk of the Galaxy (Dame et al. 2001) and in IC\,342 and Maffei\,2
(Turner \& Hurt 1992).

To derive the kinematic parameters from the $^{12}$CO(1-0) velocity
field, the method described by Begeman (1989) and implemented in the
{\sc AIPS}\ tasks ROCUR and GAL is used, although this standard
rotation curve analysis is not well suited to a galaxy as close to
face--on as NGC\,6946. The procedure fits, by least--squares, inclined
rings rotating in circular motion to the velocity field. The width of
the rings is 25\farcs0, which is slightly larger than the
$^{12}$CO(1-0) cube's beam. The initial parameter estimates are taken
from Carignan et al. (1990) and ellipse fits to the $I$--band image of
Elmegreen (2000). Points within $30\degr$ of the minor axis are
excluded (although little change in the fit results if these points
are not excluded) and a $\cos^{2}\theta$ weighting is applied.  In the
first iteration, ($\alpha_{\rm kin}$, $\delta_{\rm kin}$) marks the
position of the rotation centre of each ring and $V_{\rm sys}$,
$\alpha_{\rm kin}$, $\delta_{\rm kin}$, $V_{\rm rot}$, $i$ and
$\theta$, are solved for simultaneously.  The kinematic centre of the
molecular disk of NGC\,6946 is found to lie at $\alpha_{\rm kin}({\rm
J}2000) = 20^{\rm h}\ 34^{\rm m}\ 52\farcs7\ (\pm25\arcsec\,)$,
$\delta_{\rm kin}({\rm J}2000) = 60^{\circ}\ 09\arcmin\,\ 13\arcsec\,
(\pm 25\arcsec)$. Within the errors, this position is coincident with
the centre of the optical nucleus and the nucleus in radio
continuum. The systemic velocity $V_{\rm sys}=48\,\pm 3$\,km\,s$^{-1}$
is in excellent agreement with the the values found by other authors.

\begin{figure*}
\includegraphics*[angle=0,totalheight=8cm]{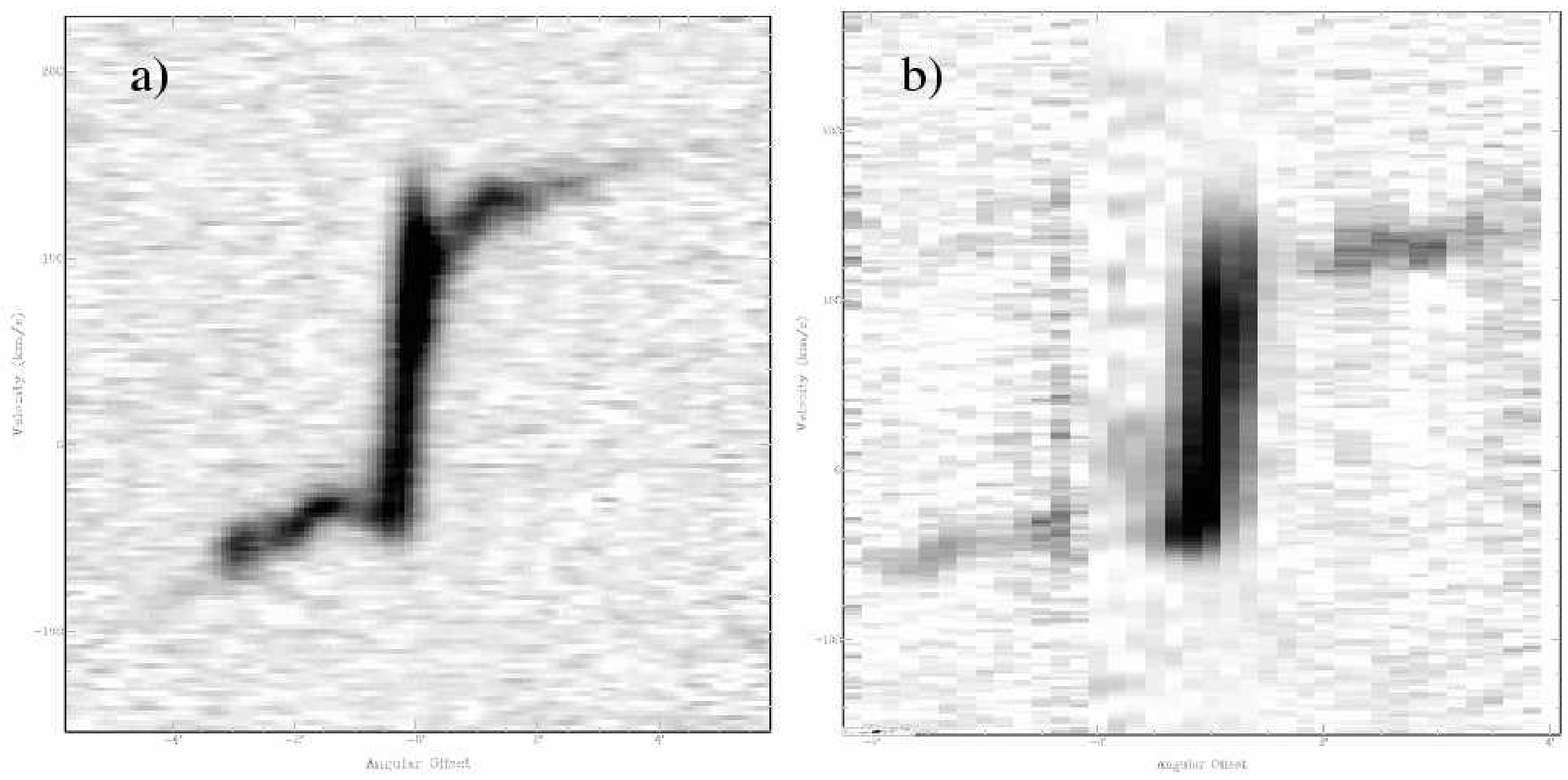}
\hfill \parbox[b]{17.5cm}{\caption[]{
\label{f:lv}Position-velocity diagrams for the $^{12}$CO(1-0) data
 (a) and the $^{12}$CO(3-2) data (b). The figures are made by taking a
 beam-wide slice through the data cubes using the position angle of
 240$\degr$ determined by Carignan et al. (1990) from {H\,{\sc i}}\
 data.
 \label{f:pv}            }    }
\end{figure*}

\begin{figure}[h]
\resizebox{\hsize}{!}{\includegraphics*[angle=0,totalheight=14cm]{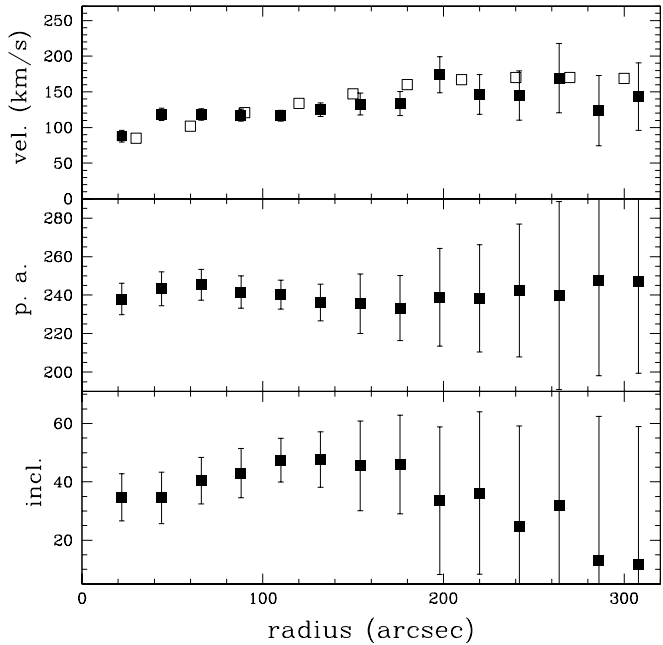}}
\caption[]{The solid squares in the upper plot shows the rotation curve 
obtained from the $^{12}$CO(1-0) velocity field, with error bars as
reported by the fitting algorithm. The {H\,{\sc i}}\ rotation curve
from Carignan et al. (1990) is shown as open symbols: the error bars
for the {H\,{\sc i}}\ data are comparable in size to the size of the
symbols. The middle panel shows the position angle as a function of
radius from the $^{12}$CO(1-0) velocity field.  The lower panel shows
the observed inclination as a function of radius. }
\label{f:rcpai}
\end{figure}

The kinematic centre and systemic velocity are well constrained and
are held fixed for every ring. Numerous fitting strategies were
attempted to estimate the variation of $i$, $\theta$ and $V_{\rm rot}$
with radius. Coupling between parameters makes a final choice of
kinematic parameters somewhat subjective for NGC\,6946. The rotation
curve shown in Fig.~\ref{f:rcpai} was determined allowing the
inclination and position angle to vary freely, and the run of these
parameters is also shown in Fig.~\ref{f:rcpai}. The best-fitting
inclination at radii less than 200\arcsec\, appears to be always
greater than the $30\degr$\, usually quoted for NGC\,6946.

The PV diagram, Fig.~\ref{f:pv}, reveals a central asymmetry in
NGC\,6946. Rapidly declining rotation of about 10--20\% of the peak
rotation velocity is evident in the inner arcminute of the galaxy,
whereas beyond this radius the rotation curve starts rising again up
to a maximum deprojected velocity of some
170$\,\pm\,10$\,km\,s$^{-1}$. There is a suggestion that at some
locations the $^{12}$CO(1-0) and the $^{12}$CO(3-2) components trace
somewhat different dynamical regimes, as the $^{12}$CO(3-2) PV
diagram, which has identical spatial and velocity resolution to the
$^{12}$CO(1-0) data, does not show the innermost decline on the
receding side, nor the kink at radius 1\farcm5 on the approaching
side. The $^{12}$CO(3-2) rotation curve appears somewhat
asymmetric. On the receding side, the velocities remain constant with
a deprojected velocity of 160$\,\pm\,10$\,km\,s$^{-1}$ while on the
approaching side the $^{12}$CO(3-2) rotation curve climbs
monotonically to a maximum deprojected velocity of
170$\,\pm\,10$\,km\,s$^{-1}$. The central ``dip'' evident in the
$^{12}$CO(1-0) PV diagram is not evident in the $^{12}$CO(3-2)
spectra, which unfortunately are very weak in the region immediately
beyond the centre at radii where the $^{12}$CO(1-0) declines. A
difference image made by subtracting the two CO velocity fields shows
now large-scale structure.

As discussed by Sofue et al. (1988) the central CO spectra, as seen in
Fig.~\ref{f:lv}, show velocities rising above the general trend of the
rotation curve. This high velocity emission may arise from a ``core''
region of molecular clouds moving in orbits not confined to the
disk. We note, however, that the rotation curve presented in
Fig.~\ref{f:rcpai} indicates that the rotation derived from our
CO(1-0) data agrees very closely with the {H\,{\sc i}}\ rotation curve
of Carignan et al. (1990) and is also very similar to that shown by
Rogstad \& Shostak (1972), but differs considerably in both shape and
amplitude from that of Sofue (1996). This difference cannot be
accounted for solely on the basis of Sofue's higher spatial resolution
(a beam-smeared model of a disk with his rotation curve is still
20\,km\,s$^{-1}$ higher than ours) and differing estimates of
NGC\,6946's kinematic parameters. This discrepancy warrants further
study. That the $^{12}$CO(1-0) rotation curve follows that of the
{H\,{\sc i}}\ supports the argument of Lavezzi \& Dickey (1998) that
CO linewidths can be indicative of the amplitude of rotation in late
type disk galaxies.

The projected velocity width integrated over one beamwidth of the
central region is 207\,km\,s$^{-1}$. From $m(r) = v^2 r/G$ we estimate a
total dynamical mass enclosed within this core of $7.7 \times
10^9~\mathrm{M}_{\odot}$, of which at least $1.3 \times
10^9~\mathrm{M}_{\odot}$ is molecular gas (see below).

The second moment images show that, in the central 2\,kpc, the average
velocity dispersion of the $^{12}$CO(3-2)  gas is $38 \pm 12$\,km\,s$^{-1}$
while for the $^{12}$CO(1-0) the dispersion is $42 \pm 7$\,km\,s$^{-1}$.
In an equivalent area located in the disk, east of the nuclear region,
the average $^{12}$CO(3-2)  gas velocity dispersion is $6.0 \pm
1.6$\,km\,s$^{-1}$ while for the $^{12}$CO(1-0) the dispersion is $8.9
\pm 2.2$\,km\,s$^{-1}$. Thus, within the $1\sigma$ uncertainties, the
velocity dispersion of the two molecular tracers is the same.  Blitz
et al. (1984) find an average velocity dispersion of
$6\pm1$\,km\,s$^{-1}$ for nearby high latitude clouds, while Stark
(1984), Stark and Brand (1989) and Malhotra (1994) find velocity
dispersions of $8\pm1$\,km\,s$^{-1}$ using differing analyses of
Galactic clouds. Malhotra (1994) finds an increase in Galactic
molecular cloud velocity dispersion with radius, which is not seen in
NGC\,6946.  Combes \& Becquaert (1997), who observe average molecular
velocity dispersions of $6$\,km\,s$^{-1}$ and $8$\,km\,s$^{-1}$ in
NGC\,628 and NGC\,3938 respectively, also do not observe radial
trends, and suggest that HI and molecular clouds are well-mixed,
forming a single kinematic gas component.  Our values for the velocity
dispersion in NGC\,6946 are thus similar to those found in the Galaxy
and in other nearby spirals.

One notable aspect of the $^{12}$CO(1-0) second moment image is that,
apart from increased dispersion in the centre, it is
featureless. Neither the spiral arms, nor the regions of high star
formation (c.f. Fig.s~\ref{f:all.data}, \ref{f:sfe}) are traced by regions
of higher dispersion. This suggests that the observed dispersion is
due to motions of cool individual clouds within the beam, rather than
thermal broadening. As is generally the case for atomic gas in spiral
galaxies, where the HI gas generally has dispersion $11\pm1$\,km\,s$^{-1}$ (e.g. Kamphuis 1993; Walsh 1997) within the bright
optical disk (and not less than $6\pm1$\,km\,s$^{-1}$ outside it), the
molecular gas appears to have a constant dispersion over
the disk of NGC\,6946. This result conflicts with the suggestion
(Sellwood \& Balbus 1999) that MHD-driven turbulence is the principal
contributor to cloud dispersions.  Sellwood \& Balbus estimate that a
$\sim 3 \mu$G field gives rise to $6$\,km\,s$^{-1}$ turbulence in
neutral clouds in the outer disk of NGC\,1058: the magnetic field in
the disk of NGC\,6946 has a peak strength $\sim 13 \mu$G and is highly
variable spatially.

\subsection{Molecular gas mass in NGC\,6946\label{s:co.mass}}

The conversion of integrated line intensity of a trace molecule like
CO to surface mass density of molecular hydrogen involves making a
series of assumptions, all of which are known in at least some cases
to be poorly justified (Morris \& Rickard 1982). The strength of the
CO line depends on the CO abundance, density, temperature and optical
depth of the line (e.g. Weiss et al. 2001). Metallicity further
influences the CO-H$_2$ conversion due to the effects of dust. More
dust results in increased efficiency of H$_2$ molecule production, and
increased UV absorption, which in turn reduces the rate of
dissociation (Verter \& Hodge 1995; Arimoto et al. 1996). Although the
principal CO heating mechanism is collisions with H$_2$ molecules,
heating by the UV radiation field can play an important role,
particularly as most of the radiation emitted by molecular clouds
comes from photodissociation regions (Hollenbach \& Tielens 1997),
which are UV-illuminated regions in molecular clouds of intermediate
density and optical depth (A$_V \leq 10$). Low energy cosmic rays can
also contribute to the heating (Glassgold \& Langer 1973).

The observed correlation between $^{12}$CO(1-0) luminosity with
molecular mass (Solomon et al. 1987; Combes 1999) has a long history
of theoretical support (e.g. Penzias et al. 1972; Mauersberger \&
Henkel 1993). Strong (1994) points out that CO emission is well
correlated with gamma rays on large scales. One line of argument
states that, if the beam-averaged cloud ensemble consists of small,
clumpy and macroturbulent objects having little radiative coupling
between them (and a low volume filling factor so that there is not
much overlap within the beam on the scale of a face-on galaxy), the CO
emission is proportional to the total number of clumps and therefore
also the total mass (Wolfire et al. 1993). This may at least be
justified for the bulk of the gas in relatively diffuse clumps,
assuming that clouds are similar from galaxy to galaxy. In galaxies
like our own where the emission comes from both diffuse gas and
molecular clouds, there does not appear to be a large variation in the
CO-to-total molecular mass conversion factor, $X_{\mathrm{CO}} =
N(\mathrm{H_2})/\mathrm{I(CO)}$, although the statistics are
small. The conversion factor is currently the subject of much
discussion. For consistency with other authors we assume the commonly
applied ``standard Galactic'' value of $2.3 \times 10^{20}
\mathrm{cm}^{-2} (\mathrm{K\,km\,s}^{-1})^{-1}$ (Strong et al. 1988),
although this value possibly overestimates the true H$_2$ mass in
NGC\,6946. This value can be compared with independently estimated
values from EGRET of $1.56 \times 10^{20}\, \mathrm{cm}^{-2}\,
(\mathrm{K\,km\,s}^{-1})^{-1}$ (Hunter et al. 1997). Integrating the
total $^{12}$CO(1-0) flux from our IRAM data cube we obtain I(CO) =
$7914 \pm 100$\,Jy\,km\,s$^{-1}$ over a solid angle of $\Delta \Omega =
4.626 \times 10^{-6}$\,sr, which gives a mean integrated emission per
beam area of $<W$(CO)$> = 15.3$\,K\,km\,s$^{-1}$. Using the relation for
the conversion of CO line intensity into H$_2$ column density given by
Strong et al. (1988), we calculate a mean column density of $3.5
\times 10^{21}~\mathrm{cm}^{-2}$. Using the above solid angle and our
adopted distance of 5.5~Mpc we obtain an estimate of the total
molecular mass of M$_{\mathrm{H}_2} = 1.13 \times 10^{10}\,
\mathrm{M}_{\odot}$.

We note that the conversion factor is in fact very likely to be
different for centre and disk, as it is well known that it differs
considerably for dense and diffuse gas. It has been suggested (Lequeux
1995; Solomon et al. 1997; Papadopolous \& Allen 2000) that CO
intensities can overestimate molecular mass in the central regions of
starburst galaxies. Therefore, using a value of $1.3 \times 10^{20}
\mathrm{cm}^{-2}\, (\mathrm{K\,km\,s}^{-1})^{-1}$ for the central
1.5\,kpc of NGC\,6946, we obtain a total molecular gas mass of $6.3
\times 10^8~\mathrm{M}_{\odot}$ for the central 1\,kpc region, and a
revised total molecular gas mass of $9.3 \times
10^9~\mathrm{M}_{\odot}$ for NGC\,6946. Valentijn et al. (1996) place
an upper limit of $3.0 \times 10^8~\mathrm{M}_{\odot}$ for the mass of
molecular gas in the central 5\arcsec\, region, and Ishizuki et
al. (1990) estimate a similar amount within the inner 300\,pc of
NGC\,6946.

Another caveat to the total gas mass determination is the fact that
optically thin CO emission has been reported in the centres of some
galaxies, on the basis of high CO(2-1)/CO(1-0) line ratios (Knapp et
al. 1980; Wall et al. 1993). Wall et al. (1993) used matched
20\arcsec\, beams to compare the first 3 J CO transitions in the
centre of NGC\,6946 and Valentijn et al. (1996) compared various
estimates of the central mass in NGC\,6946 with the result that the
H$_2$ mass within $5\farcs0$ must be in the range 2.2 -- 30 $\times 10^7\,\mathrm{M}_{\odot}$.

The optically thin thermal emission from dust grains in ``standard''
interstellar clouds is proportional to the hydrogen column density,
$N_{\mathrm{H}} = N(\mathrm{HI}) + 2N(\mathrm{H_2})$, with some
dependence on grain properties such as size, temperature and
composition (Braine et al. 1997). Thus we can compare the gas mass
obtained here with estimates from dust emission. For
example, using the dust mass estimates for NGC\,6946 from Alton et
al. (1998) listed in Table~\ref{t:general} and the dust-to-gas mass
ratios for nearby spiral galaxies (Issa et al. 1990), we estimate a
total gas mass for NGC\,6946 of 1.0 - 4.7$\times 10^{10}\,
\mathrm{M}_{\odot}$. The range of values encompasses the total gas
mass estimated above using the standard Galactic conversion factor,
which itself is in good agreement with previous estimates of the total
molecular gas in NGC\,6946 (c.f. Table~\ref{t:general}). We thus
confirm that the molecular component in NGC\,6946 is almost as massive
as the atomic gas mass, with
$\mathrm{M}_{\mathrm{H}_2}/\mathrm{M}_{\mathrm{HI}} = 0.57$. Compared
with a median $\mathrm{M}_{\mathrm{H}_2}/\mathrm{M}_{\mathrm{HI}}$
value of 0.024 for 27 Scd galaxies (Casoli et al. 1998), our result
shows that NGC\,6946 has an exceptionally massive molecular gas
component, though it is not as extreme an example as M51, with
$\mathrm{M}_{\mathrm{H}_2}/\mathrm{M}_{\mathrm{HI}} = 3.5$ (Nakai \&
Kuno 1995).

The total molecular gas mass estimated here agrees well with previous
estimates by Young \& Scoville (1982) and Tacconi \& Young
(1986). This mass represents about an order of magnitude more
molecular gas mass than is typical for spiral galaxies (Boselli et
al. 1997; Casoli et al. 1998) and is even comparable with the largest
molecular gas mass ( $1.4\times 10^{10}\,\mathrm{M}_{\odot}$) found
for a sample of interacting galaxies studied by Horellou \& Booth
(1997).

\subsection{Line ratio analysis of the ISM in NGC\,6946}

The J=4 to 1 levels of the CO molecule are 55, 33, 17 and 5.5\,K above
the ground level, respectively, and generally their emission traces
gas of corresponding excitation temperatures.  However, to reliably
decouple temperature, density and optical depth effects, measurement
of more than one of these transitions, and ideally more than one
isotopomer, is required. In the context of radiative transfer models,
temperatures of T$_k > 40$\,K and densities n(H$_2) > 3 \times
10^3$\,cm$^{-3}$ occur if the CO(3-2)/CO(1-0) line ratio is around 1.0.

\begin{figure*}[htb]
\includegraphics*[angle=0,totalheight=12cm]{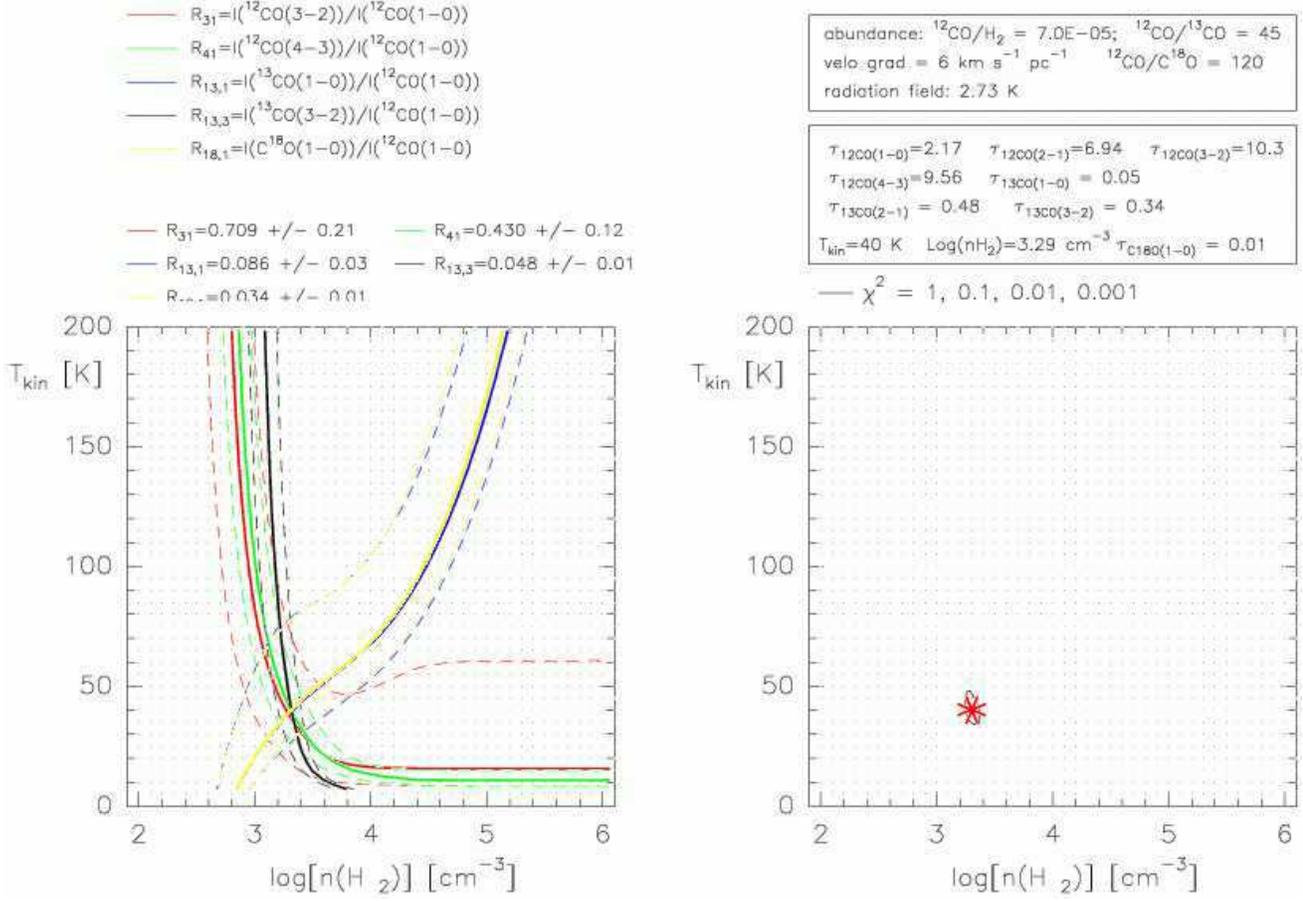}
\hfill \parbox[b]{17.5cm}{\caption[]{
Results of a LVG analysis for the central position in NGC\,6946. The
line intensity ratios have been used to derive the kinetic temperature
and density of the gas. The plot, with lines drawn for different
observed intensity ratios in various lines, emphasizes the need for
both isotopic ratios (see text). Using the ratios plotted, an excellent constraint can be placed on the temperature (40$\pm5$\,K) and
density (log(n(H$_2$))~$\sim$ 3.3\,cm$^{-3}$). }
\label{f:lvg}}
\end{figure*}

In this section, we carry out radiative transfer calculations for the
excitation of the various $^{12}$CO and $^{13}$CO transitions using
the Large Velocity Gradient (LVG) approximation (Goldreich \& Kwan
1974) with collision rates taken from Green \& Chapman (1978) to
explore non-LTE conditions. Our LVG radiative transfer program is
based on that described by Henkel (1980), and fully described in Weiss
(2000) and Weiss et al. (2001).

The model assumes a spherical, isothermal one-component ISM with a
velocity gradient sufficiently large to ensure that the source
function is locally defined (c.f. Mao et al. 2001). For optically thin
CO(1-0), the thermalization density is about 740\,cm$^{-3}$ while the
highest density for which CO intensities change significantly with
H$_2$ density is a few times $10^4$. We thus calculate LVG line
intensities for a kinetic temperature and $\mathrm{H}_2$ density range
from 5\,K to 200\,K by 5\,K and $\mathrm{log}\, n(\mathrm{H}_2)$ from
$1.8$ to $5.0$ by $0.2$, respectively. In addition, we vary the CO
abundance relative to $\mathrm{H}_2$, [CO], per velocity gradient and
the fractional $^{13}\mathrm{CO}$ and $^{12}\mathrm{C}^{18}\mathrm{O}$
abundances ([CO]/grad(V): $1 \times 10^{-5}$ to $2\times 10^{-4}$ by
$1\times 10^{-6}; \mathrm{[CO]}/{^{13}\mathrm{CO}}$: 30 to 100 by 5;
[CO]/[${^{12}\mathrm{C}^{18}\mathrm{O}}$]: 100 to 300 by 20). For the
comparison between the observed peak intensity ratios and the
predicted LVG ratios we use a $\chi ^2$ test. We use the peak values
of the spectra from the $^{12}$CO(1-0) and $^{12}$CO(3-2) data cubes
described in the previous sections, together with several pointed
observation, the results of which are listed in Table~\ref{t:spectra}.

When comparing line profiles it is important to ensure that the 
same regions in the galaxy are being probed by the different line
observations. We have mapped areas near the centre of NGC\,6946 in the
$^{12}$CO(1-0), $^{12}$CO(3-2), $^{13}$CO(3-2) and $^{12}$CO(4-3)
lines. The beam of the 30-m Pico Veleta telescope at 115\,GHz is a very
close match to the beam of the 10-m HHT at 345\,GHz, and all maps were
smoothed, using a circular Gaussian, to a common resolution of 22\arcsec. 
In Table~\ref{t:spectra} we list the beamsizes for all transitions
observed. The beamsizes have been estimated by assuming that the ratio
of the well-measured beamsizes listed in Sect.~\ref{s:observations} to
the resolution expected for these beams from $\frac{\lambda}{D}$ also
applies to nearby frequencies. Where mapped observations were not
available, we estimate a correction factor, using a standard
technique. The correction factors, $k$, listed in
Table~\ref{t:spectra} are estimated using $k = (\Theta_s^2 + \theta_1)
/(\Theta_s^2 + \theta_2)$ where $\theta_1$ is the beamsize measured,
$\theta_2$ is the beamsize with which this observation is to compared,
and $\Theta_s$ is the estimated angular extent of the source. We use
the $^{12}$CO(1-0) and $^{12}$CO(3-2)  images to estimate the angular extent of the
sources. We note that in each case the correction factors are close to
unity, and the implied corrections are of the same order as the
uncertainties in the line calibrations. For the CO(2-1) lines we have
the added check that measurements were made using both the 30-m and the
10-m telescopes.

The resulting total column density comes from the CO and
$\mathrm{H}_2$ densities, the velocity gradient and the observed line
widths using N(CO) = $3.08 \times 10^{18}\,
n(\mathrm{CO})\, \frac{dV}{\mathrm{grad(V)}}$ and N(H$_2$) = $3.08 \times
10^{18}\, n(\mathrm{H}_2)\, \frac{dV}{\mathrm{grad(V)}}$, where $dV$ is the
observed line width. A summary of the results is given in Table~\ref{t:lvg}.

\begin{table}[htb]
\caption[]{Results of the large velocity gradient model.}
\label{t:lvg}
\[
\begin{tabular*}{\linewidth}[hbt]{lcccc}
\hline
\noalign{\smallskip}
Parameter       & Value \\
\noalign{\smallskip}
\hline
\noalign{\smallskip}
\multicolumn{2}{c}{Position (0,0): best fitting LVG parameters} \\
$^{12}\mathrm{CO}/\mathrm{H}_2$          & $7.0 \times 10^{-5}$ \\
$^{12}\mathrm{CO}/^{13}\mathrm{CO}$      & $45$ \\
$^{12}\mathrm{CO}/^{12}\mathrm{C^{18}O}$ & $120$ \\
Velocity gradient                        & $6$\,km\,s$^{-1}$\,pc$^{-1}$ \\
Radiation field                          & 2.73\,K \\
$\tau _{^{12}\mathrm{CO}(1-0)}$          & 2.17                \\
$\tau _{^{12}\mathrm{CO}(2-1)}$          & 6.94                \\
$\tau _{^{12}\mathrm{CO}(3-2)}$          & 10.3                \\
$\tau _{^{12}\mathrm{CO}(4-3)}$          & 9.56                \\
$\tau _{^{13}\mathrm{CO}(1-0)}$          & 0.05                \\
$\tau _{^{13}\mathrm{CO}(2-1)}$          & 0.48                \\
$\tau _{^{13}\mathrm{CO}(3-2)}$          & 0.03                \\
$\tau _{^{12}\mathrm{C^{18}O}(1-0)}$     & 0.01                \\
T$_{\mathrm{kin}}$                       & $40$\,K              \\
Log(n(H$_2$))                            & 3.3  cm$^{-3}$     \\
X$_{\mathrm{CO}}$ & $4.4 \times 10^{19} \mathrm{cm}^{-2}(\mathrm{K\,km\,s}^{-1})^{-1}$ \\
Beam filling factor                      & 0.04 \\
\multicolumn{2}{c}{position (110,100): best fitting LVG parameters} \\
$^{12}\mathrm{CO}/\mathrm{H}_2$          & $8.0 \times 10^{-5}$ \\
$^{12}\mathrm{CO}/^{13}\mathrm{CO}$      & $90$ \\
$^{12}\mathrm{CO}/^{12}\mathrm{C^{18}O}$ & $280$ \\
Velocity gradient                        & $5$\,km\,s$^{-1}$\,pc$^{-1}$ \\
Radiation field                          & 2.73\,K \\
N(${^{12}\mathrm{CO}}$)                  & $4.8 \times 10^{17}$ cm$^{-2}$ \\
N(${^{12}\mathrm{H}_2}$)                 & $6.8 \times 10^{21}$ cm$^{-2}$ \\
$\tau _{^{12}\mathrm{CO}(1-0)}$          & 2.17                \\
$\tau _{^{12}\mathrm{CO}(2-1)}$          & 6.94                \\
$\tau _{^{12}\mathrm{CO}(3-2)}$          & $\gg 30$               \\
$\tau _{^{13}\mathrm{CO}(1-0)}$          & 0.38                \\
$\tau _{^{12}\mathrm{C^{18}O}(1-0)}$     & 0.12                \\
T$_{\mathrm{kin}}$                       & $15$\,K              \\
Log(n(H$_2$))                            & 3.0  cm$^{-3}$     \\
N(${^{12}\mathrm{CO}}$)                  & $1.0 \times 10^{17}$ cm$^{-2}$ \\
N(${^{12}\mathrm{H}_2}$)                  & $1.3 \times 10^{21}$ cm$^{-2}$ \\
X$_{\mathrm{CO}}$ & $1.1 \times 10^{20} \mathrm{cm}^{-2}(\mathrm{K\,km\,s}^{-1})^{-1}$ \\
Beam filling factor                      & 0.07 \\
\noalign{\smallskip}
\hline
\end{tabular*}
\]
\begin{list}{}{}
\item Note: the positions here are defined as offset in arcsec from the central position listed in Table~1.
\end{list}
\end{table}

The LVG model's results are listed in Table~\ref{t:lvg}, and for the
position whose offset in arcsec from the central position listed in
Table~1 is (0,0), the results are also shown in Fig.~\ref{f:lvg}. The
solid lines on the left hand plot in Fig.~\ref{f:lvg} trace those
regions of the temperature and density where observed ratios agree
with the predicted ratios, within the errors (the dotted lines). The
right hand plot shows contours of $\chi ^2$ resulting from a test of
the predicted line ratios compared with the observed ratios, for the
parameters varied. Figure~\ref{f:lvg} shows how the different line
ratios tightly constrain the temperature and density at the central
location of NGC\,6946 (c.f. Weiss 2000).

At position (0,0), the best fitting beam-averaged kinetic temperature
is 40\,K and the molecular gas density is log(n(H$_2$))=
3.3\,cm$^{-3}$, while at position (20,0), the temperature is 25\,K and
log(n(H$_2$))= 3.1\,cm$^{-3}$. At both of these positions the optical
depth of the $^{12}$CO(3-2)  line is high, with $\tau _{^{12}\mathrm{CO}(3-2)}
\sim 10$. At position (110,100), located in the north-eastern spiral arm, the
temperature is 15\,K and the density is log(n(H$_2$))=
3.0\,cm$^{-3}$. In the central position the 0.59\,kpc beam likely
averages molecular clouds over a wide range of environmental
conditions. Nevertheless, the derived values are significantly
different, with considerably higher temperatures and densities than
the position in the spiral arm. Our results support those of Wall et
al. (1993) who used a similar analysis to show that a warm (T$_K > 50
- 100$\,K) plus a cool gas component exist within a 20\arcsec\,
beamwidth at the center of NGC\,6946. Our results indicate that the
relative abundance of the warm component is much less in the spiral
arm than in the center.

The 2-D analysis (Sect.~\ref{s:twodim}) suggests that dust and molecular
gas are a similarly distributed. The resulting temperature of 40\,K for
the molecular gas can be compared with the estimated thermal
temperature of the 60-100\,$\mu$m-emitting dust of Devereux \& Young
(1993) who consider the $S_{100\mu m}/S_{60\mu m}$ flux ratio
within $6\farcm1$ to conclude that this dust, which comprises
only 10\% of the total dust mass, has a thermal temperature of
28--33\,K.

\section{Large-scale properties of NGC\,6946's disk\label{s:largescale}}

The dataset used for this part of the study is presented in
Fig.~\ref{f:all.data}, where all images have been smoothed to have the
same resolution of 22\arcsec. We have available images of NGC\,6946 in
the following wavebands, which are assumed to reflect the following
characteristic physical processes:

\begin{itemize}
\item {\bf UBVRI-band} optical light (Elmegreen et al. 2000). The $R$ and $I$ band 
images trace old stellar populations and are relatively unextincted so 
that they are reasonable probes of the bulk of the stellar mass. Since half 
the $B$-band light originates in population I stars (e.g. Young et al. 1985) younger than $2 \times 10^9$\,yr, it is indicative of SF over that period (Tacconi \& Young 1986). 
\item An optical extinction image, tracing obscuring dust from Trewhella (1998).
\item The {\bf H$\alpha$} emission (Ferguson et al. 1998) traces hydrogen ionised by young ($\leq 10^7$ yr), massive ($\geq 10\,M_{\odot}$) O and B stars 
in regions of currently active star formation (Kennicutt 1983). The considerable H$\alpha$ extinction present in NGC\,6946 (Devereux \& Young 1993) should 
be borne in mind when considering the radial surface brightness profile, 
and it has recently been claimed that SF rates determined solely from 
H$\alpha$ can underestimate the true value by 50\% (Patel \& Wilson 1995).
\item The {\bf ISOCAM ${\mathbf 7\,\mu}$m} and {\bf ${\mathbf 15\,\mu}$m} images of 
Dale et al. (1999; 2000) reflect the presence of the small, warm dust grains that are considered to be the most effective at converting incident UV flux on a molecular cloud into kinetic energy of the molecular gas (Siebenmorgen \& Kr\"ugel 1992; Hollenbach \& Tielens 1997). Another contribution also comes from PAH molecules, whose radiation does not reflect an equilibrium with the local radiation field.
\item We do not have an image of suitable spatial resolution in  {\bf far-IR} light (72\% of which comes from dust \begin{figure}[H]
\resizebox{\hsize}{!}{\includegraphics*[angle=0,totalheight=18cm]
{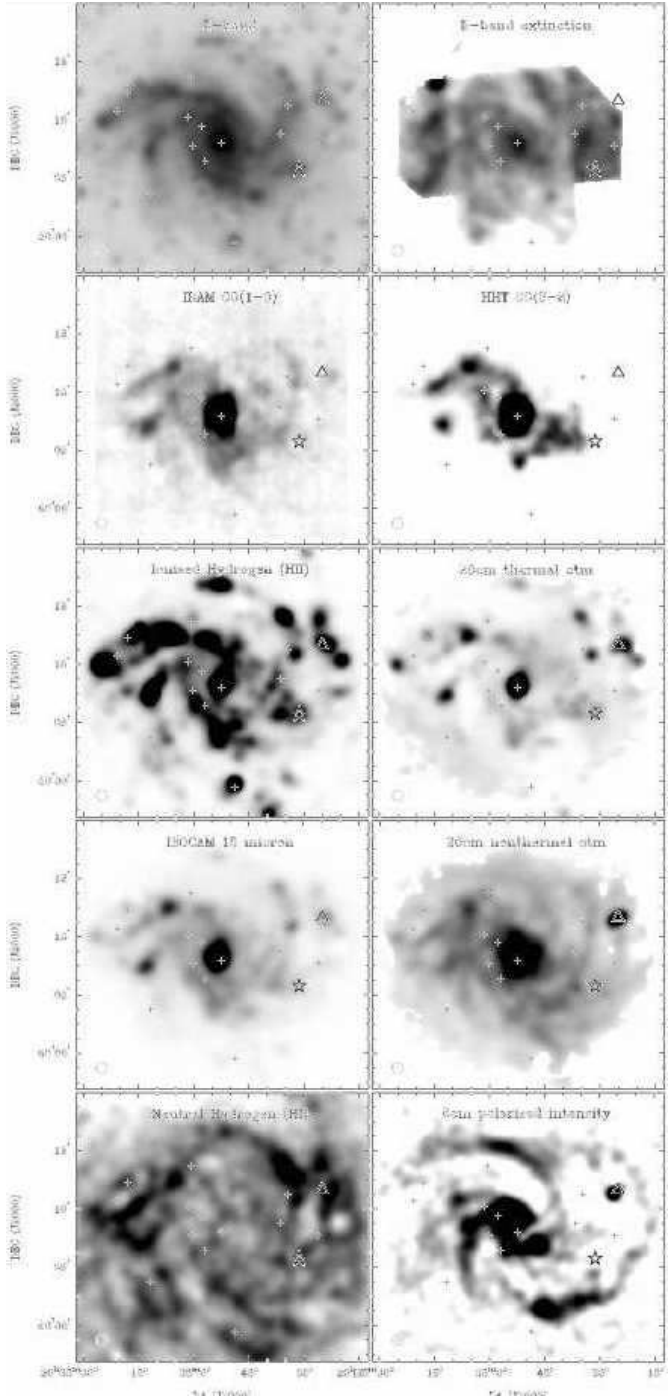}}
\caption[]{Summary of the datasets in various wavebands available 
for NGC\,6946. All images have been smoothed to have the same 22\arcsec\, 
resolution as the $^{12}$CO(1-0) data. The locations of the X-ray point 
sources discussed by Schlegel et al. (2000) are represented by small 
crosses, the large young stellar cluster identified by 
Elmegreen et al. (2000) is shown as a star and a triangle marks the 
location of a very bright, polarised radio continuum source. \label{f:all.data}}
\end{figure}
in NGC\,6946, with the rest from stars; Sauty et al. (1995)). Thus we use the radial profile presented by Devereux \& Young (1993) to trace the bulk of the dust mass.
\item The {\bf $^{12}$CO(3-2) } and {\bf $^{12}$CO(1-0)} images presented in this paper, 
representing the molecular gas.
\item The {\bf {H\,{\sc i}}} image from Kamphuis \& Sancisi (1993). 
\item The {\bf total radio continuum} $\lambda$6\,cm image (Beck \& Hoernes 1996) shows strong emission in NGC\,6946 indicating the presence of strong magnetic fields and high cosmic-ray energy density.
\item From the {\bf polarised radio} $\lambda$6\,cm image (Beck \& Hoernes 1996) we trace the regular part of the magnetic field. Magnetic field lines should follow the stream lines of the gas if the density wave compresses gas and fields in a shock front (e.g. Roberts \& Yuan 1970). However, the polarised intensity is 
concentrated {\it between} the spiral arms and may be regarded as phase-shifted images of the gas arms (Frick et al. 2000).
\item The {\bf radio spectral index} image is calculated from $\lambda$3\,cm (VLA and Effelsberg) and $\lambda$20\,cm (VLA) images (Beck, unpublished) and allows us to separate the thermal and nonthermal emission components by assuming that the average spectral index between the spiral arms of -0.95 represents that of the synchrotron radiation and that any flatter spectra are due to a thermal contribution.
\item The {\bf thermal radio continuum} $\lambda$20\,cm image traces free-free emission 
caused by the ionizing flux of O and B stars. 
\item The {\bf nonthermal radio continuum} $\lambda$20\,cm image, which shows the synchrotron radiation from relativistic electrons in NGC\,6946's magnetic field, may be expected to be correlated with molecular gas as the supernova remnants embedded in the ISM are the sources of cosmic-ray particles. However, in case of (often assumed) equipartition between the energy densities of cosmic rays and magnetic fields, the field strength is the dominant parameter for synchrotron intensity (see Sect.~\ref{mir.radio}).
\end{itemize}

\subsection{One-dimensional properties of the disk\label{s:onedim}}

The radial surface brightness profile of the optical light of a spiral
galaxy can generally be separated into a central spheroid and a
flattened disk component. The optical disk can almost always be well
fitted with a function of the form $I(r) = I_0 \mathrm{exp}(-r/h)$
where $I(r)$ is the intensity at radius $r$, $I_0$ is the intensity at
$r = 0$ and $h$ is the exponential scale length. Such a fitting
function is often a good match to azimuthally averaged profiles of the
atomic and molecular gas as well. Knapen \& van der Kruit (1991) point
out that different authors often report significantly different scale
lengths for the same galaxy and stress the care that must be taken to
obtain a reliable value. To ensure that fitting errors are minimal, we
attempt to obtain the brightness profiles directly from the data in a
uniform manner.

\begin{figure}[bht]
\resizebox{\hsize}{!}{\includegraphics*[angle=0,totalheight=9cm]{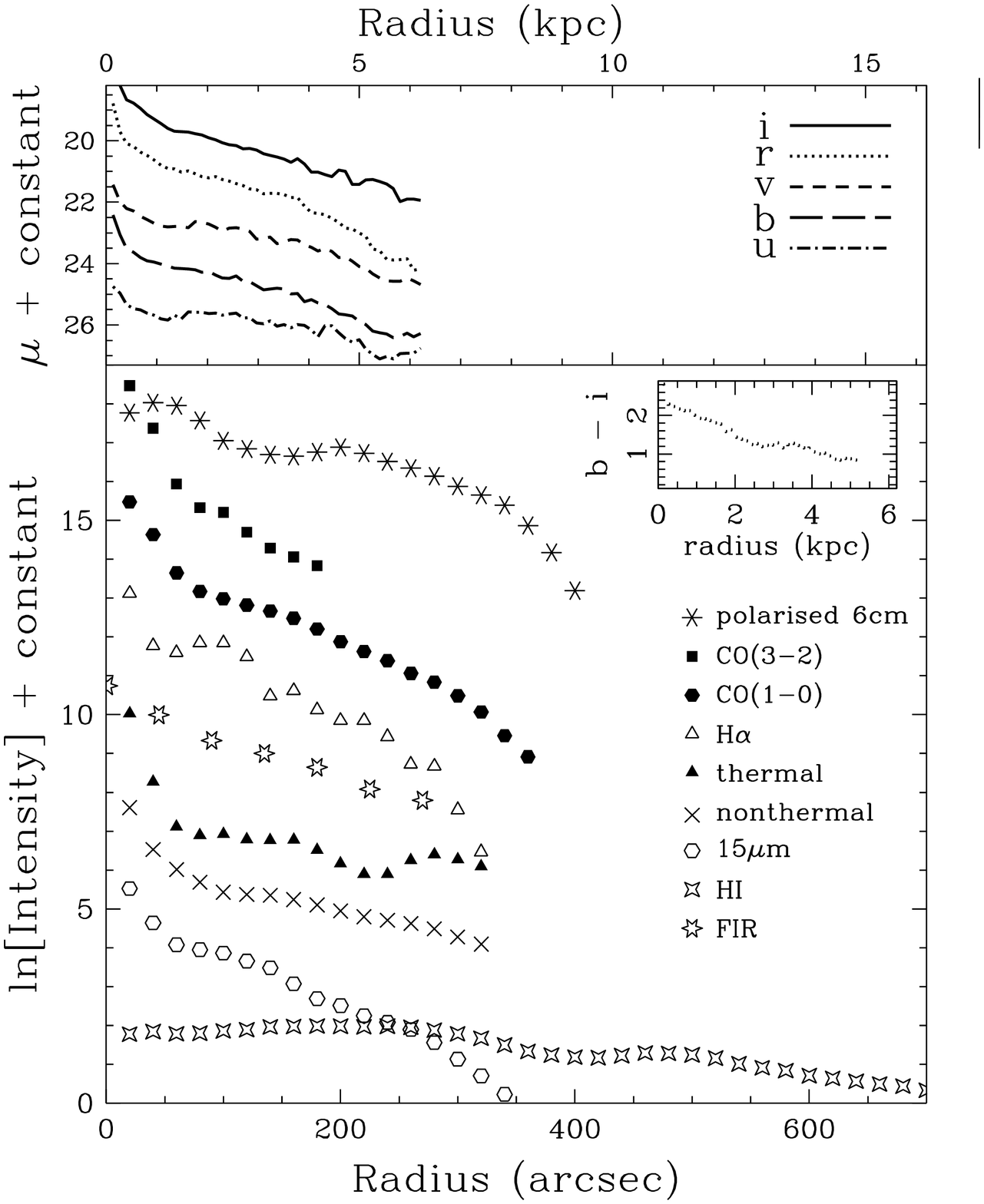}} 
\caption[]{Azimuthally-averaged radial trends of various tracers in NGC\,6946. 
The optical data is that of Larsen \& Richter (1999). The Galactic extinction 
corrected $V$-band data is correctly scaled - the profiles for the other bands 
are offset for visibility. The log values of other tracers are plotted on an 
arbitrary scale simply to show the differing radial trends. All plots were 
made assuming the kinematic orientation parameters derived from the 
{H\,{\sc i}}\ velocity field by Carignan et al. (1990) who find 
i = $38\degr$ and p.a. = $240\degr$. \label{f:allvr}}
\end{figure}

The azimuthally-averaged radial distributions of various tracers are
presented in Fig.~\ref{f:allvr} and the results of least squares
fitting to the profiles are listed in Table~\ref{t:rsbp}. All profiles
are obtained by deprojecting the galaxy assuming an inclination of
$38\degr$ and a position angle of $240\degr$ (Carignan et al. 1990).

The fitting is done only to the region beyond the ``nuclear
concentration'', where a second exponential disk with a much steeper
profile is more appropriate. Sofue et al. (1988) assert that this
nuclear concentration is a structure distinct from the extended
disk. The FIR points in Fig.~\ref{f:allvr} are taken from Table 2 of
Devereux \& Young (1993). The optical, H$\alpha$, MIR and {H\,{\sc
i}}\ data are from Larsen \& Richter (1999), Ferguson et al. (1998),
Dale et al. (2000) and Kamphuis \& Sancisi (1993), respectively,
kindly provided by those authors. The radio continuum data are
described in more detail in Beck (1991) and Beck \& Hoernes (1996).

\begin{table}[htb]
\caption[]{Exponential-disk scale lengths}
\label{t:rsbp}
\[
\begin{tabular*}{\linewidth}[hbt]{lcccc}
\hline
\noalign{\smallskip}
Waveband       & Range of fit&\multicolumn{2}{c}{Scale length} & Source$^{\mathrm{a}}$ \\
               &   (\arcsec)  &   (\arcsec)   & (kpc)            &             \\
\noalign{\smallskip}
\hline
\noalign{\smallskip}
$U$            & 70-200 & 92$\pm18$ & 2.5 & 1 \\
$B$            & 70-200 & 92$\pm18$ & 2.5 & 1 \\
$V$            & 70-200 & 81$\pm16$ & 2.2 & 1 \\
$R$            & 70-200 & 80$\pm16$ & 2.2 & 1 \\
$I$            & 70-200 & 82$\pm16$ & 2.2 & 1 \\
$R$  DSS       & 70-200 & 93$\pm18$ & 3.4 & 2 \\
$R$  Palomar   & 70-200 &116$\pm23$ & 3.1 & 3 \\
$R$  Lowell    & 70-200 & 86$\pm17$ & 2.3 & 4 \\
H$\alpha$      & 70-200 & 72$\pm14$ & 1.9 & 3 \\
H$\alpha$      & 70-200 & 82$\pm16$ & 2.2 & 1 \\
H$\alpha$      & 70-200 & 64$\pm13$ & 1.7 & 5 \\
H$\alpha$      & 60-260 & 71$\pm14$ & 1.9 & 5 \\
6\,cm P.I.      & 60-260 &169$\pm30$ & 4.6 & 6           \\
20\,cm therm.   & 60-260 &157$\pm30$ & 4.3 & 6           \\
20\,cm nonth.   & 60-260 &146$\pm27$ & 4.0 & 6           \\
CO(1-0)        & 60-260 & 83$\pm16$ & 2.2 & 6           \\
CO(3-2)        & 60-180 & 57$\pm12$ & 1.5 & 6           \\
{H\,{\sc i}}   &200-600 &324$\pm40$ & 8.7 & 7 \\
15\,$\mu$m      & 60-260 & 84$\pm16$ & 2.2 & 8 \\
FIR            & 80-270 & 80$\pm16$ & 2.2 & 9 \\
\noalign{\smallskip}
\hline
\end{tabular*}
\]
\begin{list}{}{}
\item[$^{\mathrm{a}}$] 1 = Larsen \& Richter (1999); 2 = Lasker et al. (1990); 
3 =  Malhotra et al. (1996); 4 = Dettmar (priv. comm.); 5 = Ferguson et al. (1998); 
6 = this paper; 7 = Kamphuis \& Sancisi (1993); 8 = Dale et al. (1999); 
9 = Devereux \& Young (1993)
\end{list}
\end{table}

The optical data from Larsen \& Richter (1999) do not cover the entire
disk of NGC\,6946 so the radial profiles are only reliable out to a
radius of about 200\arcsec. Nevertheless, from Fig.~\ref{f:allvr} and
Table~\ref{t:rsbp} it can be seen that there are significant
differences between the optical scale lengths within the inner few kpc
of the disk, where the relative photometry is likely to be
reliable. The inset in Fig.~\ref{f:allvr} shows the $B - I$ colour
gradient to be strongly negative, as expected for a dusty, late type
galaxy (Wirth \& Shaw 1983; Balcells \& Peletier 1994). This colour
gradient is also dramatically evident in Fig. 1 of Elmegreen et
al. (2000), a true-colour representation of NGC\,6946 using this same
optical data. Baffa et al. (1990) have previously found
evidence for strong NIR-optical colour gradients in NGC\,6946, which
they ascribe to intrinsic reddening. A $B - I$ image (e.g. Trewhella
1998) shows the considerable Galactic extinction in NGC\,6946 (A$_U =
1.858$, A$_B = 1.475$, A$_V = 1.133$, A$_R = 0.914$, A$_I = 0.663$:
Schlegel et al. (1998)), which makes it difficult to interpret the colours of
NGC\,6946's components in terms of stellar populations. 

Ryder and Dopita (1994) find that, for a sample of 34 spiral galaxies,
the H$\alpha$ scale length in the outer disk is significantly longer
than the $I$ scale length. This is not the case in NGC\,6946, where,
within the uncertainties, the scale lengths are equal, and possibly
the H$\alpha$ scale length is marginally shorter than the optical
values if the Ferguson et al. (1998) dataset is considered. 

The bright, starbursting centre of NGC\,6946 (Engelbracht 1996) can be
seen in the rapidly rising optical radial surface brightness profiles,
and, within a radius of 1.5\,kpc, the $^{12}$CO(1-0), $^{12}$CO(3-2) ,
H$\alpha$, total $\lambda$6\,cm, $\lambda$20\,cm thermal and nonthermal
radio continuum emission and the 15\,$\mu$m profiles also show a steep
rise.

\begin{table*}[h]
\caption[]{Correlations $X\longleftrightarrow Y$ between the logarithms of 
the intensities of various emissions in NGC\,6946, computed between 0.5 and 
6\arcmin\, distance from the centre in the maps at 22\arcsec\, resolution.
The first value tabulated is the slope of the regression line (and its error),
followed by the Pearson's correlation coefficient $r$ (and its error) and 
the $t$ value of Student's t-test (see text for details).}
\label{t:corr}
\[
\begin{tabular*}{\linewidth}[tbh]{lllllllllll}
\hline
\hline
\noalign{\smallskip}
$X\,\,\,\,\,\setminus\,\,\,\,\,Y$ & H$\alpha$ &7\,$\mu$m  &15\,$\mu$m &6\,cm  &Thermal &Nonthermal &6\,cm PI   &{H\,{\sc i}}\ &CO(1-0)$^{\mathrm{a}}$ &Total gas \\
\noalign{\smallskip}
\hline
\hline
\noalign{\smallskip}
R-band                & 1.46      & 1.26       & 1.29      & 0.88    &$\approx$0.8 & 0.84  &$\approx$0.6   & --        & 1.3       & 0.5       \\
                      & $\pm$0.05 & $\pm$0.09  & $\pm$0.22 & $\pm$0.06 & --        & $\pm$0.07 & --        & --        & $\pm$0.6  & $\pm$1.2  \\
                      & 0.84     & {\bf 0.92}  & 0.90      & 0.90      & 0.61      & 0.81      & 0.43      & $<$0.1    & 0.75      & 0.50      \\
             & $\pm$0.02     & {\bf$\pm$0.01}  & $\pm$0.02 & $\pm$0.01 & $\pm$0.07 & $\pm$0.03 & $\pm$0.08 & --        & $\pm$0.05 & $\pm$0.07 \\
                      & 19       & {\bf 27}    & 20        & 27        & 6.7       & 15        & 4.4       & $<$1      & 9.8       & 6.3       \\
\hline
\noalign{\smallskip}
H$\alpha$             & xxx       & 0.83       & 0.79      & 0.63      & 0.8       & 0.60      & --      &$\approx$0.3 &$\approx$0.8 & 0.4     \\
                      &           & $\pm$0.06  & $\pm$0.18 & $\pm$0.08 & $\pm$0.6  & $\pm$0.14 & --        & --        & --        & $\pm$0.5  \\
                      & xxx       & 0.86       & 0.81      & 0.82      & 0.72      & 0.62      &$<$0.1     & 0.40      & 0.58      & 0.65      \\
                      &           & $\pm$0.02  & $\pm$0.03 & $\pm$0.03 & $\pm$0.06 & $\pm$0.06 & --        & $\pm$0.08 & $\pm$0.07 & $\pm$0.05 \\
                      & xxx       & 19         & 14        & 17        & 8.6       & 8.5       &$<$2       & 4.5       & 6.4       & 9.1       \\
\hline
\noalign{\smallskip}
7\,$\mu$m              &           & xxx        & 1.02      & 0.75      & 0.8       & 0.69   &$\approx$0.5 &$\approx$0.4 & 1.2      & 0.51      \\
                      &           &            & $\pm$0.02 & $\pm$0.06 & $\pm$0.5  & $\pm$0.07 & --        & --        & $\pm$0.3  & $\pm$0.17 \\
                      &           & xxx   &{\bf 0.983} & {\bf 0.92}    & 0.70      & 0.83      & 0.34      & 0.22      & 0.83      & 0.62      \\
                      &           & &{\bf $\pm$0.003}  &{\bf$\pm$0.01} & $\pm$0.06 & $\pm$0.03 & $\pm$0.09 & $\pm$0.10 & $\pm$0.04 & $\pm$0.06 \\
                      &           & xxx   & {\bf 55}   & {\bf 28}      & 8.4       & 16        & 3.4       & 2.2       & 13        & 8.2       \\
\hline
\noalign{\smallskip}
15\,$\mu$m               &         &            & xxx       & 0.77      & 0.9       & 0.71   &$\approx$0.6  & --        & 1.19    & 0.58      \\
                        &         &            &           & $\pm$0.10 & $\pm$0.5  & $\pm$0.11 & --        & --        & $\pm$0.22 & $\pm$0.16 \\
                        &         &            & xxx       & 0.90      & 0.70      & 0.80      & 0.34      &$<$0.1     & 0.85      & 0.58      \\
                        &         &            &           & $\pm$0.02 & $\pm$0.06 & $\pm$0.04 & $\pm$0.10 & --        & $\pm$0.03 & $\pm$0.07 \\
                        &         &            & xxx       & 21        & 8.1       & 13        & 2.8       &$<$1       & 14        & 7.0       \\
\hline
\noalign{\smallskip}
6\,cm                    &         &            &           & xxx       & 0.9       & 0.91   &$\approx$0.6 &$\approx$0.5 & 1.3      & 0.6       \\
                        &         &            &           &           & $\pm$0.4  & $\pm$0.07 & --        & --        & $\pm$0.3  & $\pm$0.3  \\
                        &         &            &           & xxx       & 0.73      & 0.92      & 0.52      & 0.28      & 0.80      & 0.66      \\
                        &         &            &           &           & $\pm$0.05 & $\pm$0.02 & $\pm$0.07 & $\pm$0.09 & $\pm$0.04 & $\pm$0.05 \\
                        &         &            &           & xxx       & 9.4       & 25        & 6.4       & 3.0       & 12        & 9.6       \\
\hline
\noalign{\smallskip}
Thermal                 &         &            &           &           & xxx       & 1.2       & --      &$\approx$0.6 &$\approx$1.2 & 0.7       \\
                        &         &            &           &           &           & $\pm$0.3  & --        & --        & --        & $\pm$0.4  \\
                        &         &            &           &           & xxx       & 0.53      & $<$0.2    & 0.22      & 0.48      & 0.53      \\
                        &         &            &           &           &           & $\pm$0.08 & --        & $\pm$0.12 & $\pm$0.10 & $\pm$0.08 \\
                        &         &            &           &           & xxx       & 5.5       & $<$2      & 1.9       & 4.1       & 5.4       \\
\hline
\noalign{\smallskip}
Nonthermal              &         &            &           &           &           & xxx    &$\approx$0.7  & --        & 1.5      & $\approx$0.6 \\
                        &         &            &           &           &           &           & --        & --        & $\pm$0.5  & --        \\
                        &         &            &           &           &           & xxx       & 0.59      &$<$0.1     & 0.80      & 0.43      \\
                        &         &            &           &           &           &           & $\pm$0.07 & --        & $\pm$0.04 & $\pm$0.08 \\
                        &         &            &           &           &           & xxx       & 6.8       &$<$1       & 12        & 4.9       \\
\hline
\noalign{\smallskip}
 6\,cm PI                &         &            &           &           &           &           & xxx       & --        & 1.4       & --       \\
                        &         &            &           &           &           &           &           & --        & $\pm$0.8  & --       \\
                        &         &            &           &           &           &           & xxx       & $<$0.1    & 0.46      & $<$0.2   \\
                        &         &            &           &           &           &           &           & --        & $\pm$0.10 & --       \\
                        &         &            &           &           &           &           & xxx       & $<$1      & 11        & $<$2     \\
\hline
\noalign{\smallskip}
{H\,{\sc i}}            &         &            &           &           &           &           &           & xxx       & --        & 1.2       \\
                        &         &            &           &           &           &           &           &           & --        & $\pm$0.3  \\
                        &         &            &           &           &           &           &           & xxx       &$<$0.1     & 0.76      \\
                        &         &            &           &           &           &           &           &           & --        & $\pm$0.04 \\
                        &         &            &           &           &           &           &           & xxx       &$<$1       & 12        \\
\hline
\noalign{\smallskip}
CO(1-0)                 &         &            &           &           &           &           &           &           & xxx       & 0.57      \\
                        &         &            &           &           &           &           &           &           &           & $\pm$0.15 \\
                        &         &            &           &           &           &           &           &           & xxx       & 0.61      \\
                        &         &            &           &           &           &           &           &           &           & $\pm$0.07 \\
                        &         &            &           &           &           &           &           &           & xxx       & 6.6       \\
\noalign{\smallskip}
\hline
\hline
\end{tabular*}
\]
\begin{list}{}{}
\item[$^{\mathrm{a}}$] The $^{12}$CO(3-2)  image has too few independent data points for
reliable correlations.
\end{list}
\end{table*}

The {H\,{\sc i}}\ profile, as noted by Rogstad \& Shostak (1972),
shows a slight decline towards the centre, as does the polarised
intensity profile. The {H\,{\sc i}}\ in NGC\,6946 was noted by
Tacconi \& Young (1986) to differ from other tracers: it does not
clearly follow spiral arms, unlike many other spiral galaxies for
which interferometric {H\,{\sc i}}\ data are available. The inner four
spiral arms are not distinguishable at all and the inner {H\,{\sc i}}\
disk is quite irregular. Some H$\alpha$ hotspots are also {H\,{\sc
i}}\ concentrations, but not all, and there are many regions with
H$\alpha$ but where the {H\,{\sc i}}\ is very weak. Although the very
faint H$\alpha$ arm in the north does have a clear {H\,{\sc i}}\
counterpart, there is a long arm segment in the SW that is not
discussed further here as it is located at a radius beyond all the
other tracers' limits.

The $^{12}$CO(3-2)  scale length is the shortest measured, since the warm gas
leading to emission of the $^{12}$CO(3-2)  line is commonest in the densest
and most active regions, which are concentrated towards the
centre. The $^{12}$CO(3-2)  profile in Fig.~\ref{f:allvr} appears, within the
uncertainties, to follow the H$\alpha$, both in the approximate slope
and shape (central steepening and slight bump at 100\arcsec).

The most striking aspect of Fig.~\ref{f:allvr} is that, although all
profiles are clearly not well represented by single exponentials, and
differ amongst one another greatly, the VRI, H$\alpha$, 15\,$\mu$m,
$^{12}$CO(1-0) and FIR scale lengths over several kpc are the same to
within 5\%. Close correspondence between some of these profiles has
been noted before by deGioia-Eastwood et al. (1984), Tacconi\, \&
Young (1986), Sofue et al. (1988), Devereux \& Young (1993) and
Malhotra et al. (1996).

The scale length of the thermal radio emission in Table 5 is very
large, much larger than that of the H$\alpha$ emission. Absorption in
the H$\alpha$ line is stronger in the inner region, so the H$\alpha$
scale length in Table 5 is an upper limit, making the difference to
the thermal radio emission even larger. We propose as an explanation
an artifact of the thermal-nonthermal separation.  According to
Fig.~10, the radial profile of the thermal radio emission becomes flat
in its outer part. Here, observational uncertainties in the baselevels
and the spectral indices may cause a systematic overestimate of the
thermal emission.

\subsection{Two-dimensional correlation analysis of the disk\label{s:twodim}}

Combining an image into a line and then further reducing the line to a
single value - the scale length - clearly discards much
information. In this section we attempt to use the full 2-D
distributions at various wavelengths to investigate, by correlating
images, the relationship between the ISM's components, the morphology,
magnetic field and the SF in NGC\,6946's disk.

The correlations are determined using an algorithm that measures the
logarithm of the intensity in the same spatial region in two images
(Nieten 2001). The correlation is determined using a variety of
different gridding and weighting schemes. The results presented here
use 1/$\sigma$ weighting, where $\sigma$ is the rms noise in each
image. Four different hexagonal grids are used for each correlation to
assess the effect of gridding on the resulting correlations, and a
mean value adopted. The difference in the correlation coefficient for
orthogonal or slightly offset hexagonal grids is typically a few
percent. Only spatially independent points are selected. The Gaussian
nature of the radio telescope beams means that the definition of
``adjacent independent data points'' is not well defined. Simulations
using artificial data indicate that selecting regions separated by
1.98 beamwidths, at which point the Gaussians overlap by only 2\% in
one dimension, is a good compromise between independence and maximal
data usage.

Care is taken to ensure that accurate zero levels are subtracted from
each data point before the correlation is calculated.  To reduce the
effect of errors due to baselevel uncertainties and noise
fluctuations, only datapoints with amplitudes of five times the rms
noise $\sigma$ are used. Thus both the magnitude of the resulting
correlation and also the slope of the regression line can in principle
contain physical information.

Since the physical processes associated with the central starburst are
likely to be of a fundamentally different character to those occurring
in the disk (Helfer \& Blitz 1997), we excluded the points in the
innermost (deprojected) arcminute of NGC\,6946. For most images this
had the effect of excising the few highest intensity values, which are
few in number. For the 6cm images, a strong background source on the
western side of the galaxy is removed before correlating.

The Pearson $r$ value measures a relation between two variables only
to the extent to which it is linear. Since the most common
relationships between variables of astrophysical interest is a power
law, we correlated the logarithm of the intensities rather than the
raw values. We then created a correlation matrix between all observed
variables and inspected the values for expected and unexpected
relations between variables, given the relative quality of the
data. The results of this analysis is presented in Table~\ref{t:corr}.
The slope of the regression line, the (Pearson's $r$) correlation
coefficient and the Student's t value are listed. $r^2$ is the
coefficient of determination, which measures the strength of the
relationship between the two variables, while $t$ is a measure of the
significance of the correlation.

The significance of the correlations is not well determined. One
problem is that, due to the effect of noise making smaller values less
reliable, in no case are the errors (deviations from the regression
line) normally distributed. Thus our measure of the significance of
the correlations is not well justified since the variability of the
dependent variables is not normally distributed, nor is the
variability the same for all values of the dependent
variable. However, this problem becomes less important as more
independent data points are included in the correlation. When the
number of points exceeds 50, Monte Carlo simulations suggest that the
error distribution does not introduce serious biases, and when the
number exceeds 100, the non-normal error distribution does not
significantly affect the result of the correlation analysis.  We
considered the effect of outliers on the value of the correlations by
inspecting the scatter plots individually. Since most of the images
are of high quality over the area common to all of them or are heavily
spatially smoothed to the resolution of the CO data, the above
condition is easily met for all images, except the $^{12}$CO(3-2) data
which has a limited spatial extent. Each correlation is the result of
between 95 and 312 independent data points. As the reliability of the
correlation coefficient increases with its absolute value, relatively
small differences between large correlation coefficients can be
significant. As a guideline, well-correlated samples in our dataset
have $t$ values of 12 or higher.

The strongest correlations, all of which have high statistical
significance, are: 7\,$\mu$m--6\,cm total power, 7\,$\mu$m--R-band,
6\,cm--nonthermal, R-band--6\,cm, 15\,$\mu$m--6\,cm total power,
15\,$\mu$m--R-band, 7\,$\mu$m--H$\alpha$ and
7,15\,$\mu$m--CO(1-0). The two MIR images are also significantly
correlated with the nonthermal emission, and the $^{12}$CO(1-0) correlates
with the $\lambda$6\,cm total power and the nonthermal emission. The weakest
correlations are those involving {H\,{\sc i}}\, or $^{12}$CO(3-2) . In
particular, the {H\,{\sc i}}\, image does not correlate with $^{12}$CO(1-0),
and both thermal emission and H$\alpha$ correlate very weakly with
CO(1-0). The $\lambda$6\,cm polarised intensity also shows no
significant correlation with any other image. In fact, as noted by
Frick et al. (2001), the polarised intensity is, on some spatial
scales, {\it anticorrelated} with other tracers. In particular, the
spiral pattern in the polarised intensity falls almost exactly between
the optical spiral arms.

\section{Discussion}
\subsection{Correlations within exponential disks}

A trivial observation that arises from correlation studies of spiral
galaxies, demonstrated in Fig.~\ref{f:allvr}, is that all the observed
material components that radiate have an approximately exponential
decline with radius (Hoernes et al. 1998). A purely structureless disk
will engender a strong correlation between any two tracers, so the
extent to which small scale structures fail to correlate will be
reflected in decreased correlations. To quantify the effect of, for
example, correlations arising between various disk tracers due to the
exponential disk scale length being similar, more sophisticated
correlation techniques are required. Frick et al. (2001) used a
wavelet correlation method to investigate spatial structures in
NGC\,6946 and find strong similarities between the total radio
emission, red light and mid-infrared structures on all wavelet scales.

The high correlations between R-band, CO, MIR, $\lambda$6\,cm total and
$\lambda$20\,cm nonthermal intensities are partly due to the bright
exponential disks which contain a significant fraction of the total
flux. In H$\alpha$, {H\,{\sc i}}, $\lambda$20\,cm thermal and
$\lambda$6\,cm polarised intensity, on the other hand, much of the
spatial power in NGC\,6946 is concentrated in the spiral arms rather
than in the disk so that our correlations refer to relatively small
scales. 

In M31 and in the Galaxy, the gas and dust column densities are found
to be well correlated (Savage \& Mathis 1979; Xu \& Helou 1996). Xu \&
Helou (1996) find that, in M31, the {H\,{\sc i}}\ is correlated with
the dust, which is not observed in NGC\,6946, where the only
significant correlation of {H\,{\sc i}}\ is with H$\alpha$. This is
because the {H\,{\sc i}}\ distribution in NGC\,6946 is essentially
featureless within the optical disk, but the two outer {H\,{\sc i}}\
arms have counterparts in the wide-field H$\alpha$ image.

The contrast between the {H\,{\sc i}}\ and $^{12}$CO(1-0) disks in
NGC\,6946 is revealing. A small depression in the radial {H\,{\sc i}}\
profile is common for spiral galaxies, but in NGC\,6946 the slope of
the {H\,{\sc i}}\ profile is positive out to a radius of about
190\arcsec\, or some 5\,kpc (Fig.~\ref{f:allvr} and Fig.~10 of
Boulanger \& Viallefond (1992)). The {H\,{\sc i}}\ disk correlates
poorly with both the main molecular gas tracer and with the SF tracers
H$\alpha$.  The spatial relationship between molecular clouds and
{H\,{\sc ii}}\ regions is also of interest. Solomon et al. (1985) find
that 25\% of warm molecular clouds in the Galaxy are associated with
{H\,{\sc ii}}\ regions. Averaged over the beam area, a similar
association in NGC\,6946 would lead to a strong correlation between
the H$\alpha$ and $^{12}$CO(1-0) images. However, these two images,
although somewhat correlated ($r = 0.58, t = 6.4$), are much less so
than the $^{12}$CO(1-0) is with the nonthermal radio continuum or H$\alpha$
is with the 15\,$\mu$m. Similarly, the $^{12}$CO(1-0) is relatively poorly
correlated with the $\lambda$20\,cm thermal radio continuum.

\subsection{The mid-infrared -- radio continuum correlation}
\label{mir.radio}

That the 15\,$\mu$m emission is well correlated with the molecular gas
and H$\alpha$ is unsurprising, indicating that the bulk of the
15\,$\mu$m emission comes from warm dust grains and PAH molecules in
NGC\,6946, and that these particles are in turn concentrated near
regions of current star formation. We find here, as Vogler et
al. (2001) find for M\,83, that the mid-infrared (MIR) images in the
7\,$\mu$m and 15\,$\mu$m bands are highly similar and most probably
dominated by PAHs, except in the central region where the strong
radiation field destroys some fraction of the PAHs. Vogler also found
that the 7\,$\mu$m and 15\,$\mu$m bands are most strongly correlated
with the total radio continuum. However, as no separation of thermal
and nonthermal radio continuum components was possible in M83, Vogler
et al. (2001) could not further investigate the origin of their
correlations. The weakness of the correlation between H$\alpha$ and
the thermal radio continuum may be a result of strongly varying
absorption, not reflected in the mid infrared images, which better
trace PAHs than dust temperature.

Table~\ref{t:corr} shows red light to be highly correlated with
7\,$\mu$m. If the latter is indeed due primarily to PAHs, we may
conclude that PAHs are heated by the general radiation field.

NGC\,6946 is one of the few nearby spiral galaxies for which separate
images of the thermal and nonthermal radio continuum components are
available. If the radio -- MIR correlation would be simply due to the
radiation field near star-forming regions, heating the PAHs and
ionising the gas, we would expect the best correlation between the MIR
and the thermal radio intensities.  Table~\ref{t:corr} tell us that
this is not the case: the thermal emission is {\it less} correlated
with the MIR emission than the nonthermal emission. According to Table
6, the best correlation is between MIR and {\it total} radio
continuum. This indicates that there are in fact two relationships
between radio continuum and MIR emission, one for each component,
clumpy and diffuse. The diffuse radio component is mainly nonthermal
(a thick disk, if observed edge-on), while the clumpy radio component
is mainly thermal emission from HII regions in a thin disk. While the
physical relationship between the clumpy components is obvious, the
other relationship, between the diffuse MIR component and the
synchrotron radio component, is much more difficult to explain. The
physical background of this correlation must be nonthermal. The
assumption of enhanced production of cosmic-ray particles in
star-forming regions is a possible explanation only if the magnetic
field strength is constant everywhere in NGC\,6946.  As the intensity
of synchrotron emission depends only on the first power of the
cosmic-ray density, but on a power between two and four of the field
strength, depending how well the energy equipartition condition is
fulfilled, even small fluctuations in field strength will dominate the
intensity distribution.

We propose a close connection between magnetic fields and gas clouds
as the physical background of the radio -- MIR correlation,
as already suggested for the radio -- far-infrared correlation by
Niklas \& Beck (1997) and Hoernes et al. (1998). In this case we also
expect a high correlation between the nonthermal intensity and the
total gas density.

Such a correlation was found in M\,31 (Hoernes et al. 1998), but is much
less convincing in NGC\,6946. The correlation between nonthermal radio
emission and $^{12}$CO(1-0) is reasonable, but the correlation with the total
gas is not significant (Table 6).

We conclude that the magnetic fields in NGC\,6946 are {\it not} closely
related to the neutral gas as observed in the $^{12}$CO(1-0) or HI lines, but
to the gas in which the PAHs are embedded, the {\bf warm gas}.  This
gas needs heating by the general radiation field and has a
sufficiently high degree of ionisation to couple the field lines.
This can be tested by comparing the nonthermal radio emission with
tracers of the warm gas, e.g. C+ or CO transitions higher than (3-2).

\subsection{Gas, star formation the role of the magnetic field in NGC\,6946}
\label{gas.sf}

The absence of association between {H\,{\sc i}}\ and H$\alpha$ taken
together with the considerable molecular gas reservoir throughout the
disk argues against the hypothesis (e.g. Allen et al. 1986; Smith et
al. 2000) that {H\,{\sc i}}\ is always the product of star formation,
in the sense that H$_2$ is photodissociated by UV radiation from young
stars. If {H\,{\sc i}}\ was produced in this way, then one would
expect an association of {H\,{\sc i}}\ with the causes of the
photodissociation. All correlations with a total gas mass image (made
from summing the atomic and molecular contributions) are worse than
with CO alone. Thus the situation is different from that in M31 where
the correlation between radio continuum and FIR is better with total
gas mass than with CO or HI alone (Hoernes 1997). This reflects the
unusually high ratio of molecular to atomic gas in NGC\,6946.

Compared with other nearby galaxies of comparable morphological type,
luminosity and central surface brightness for which appropriate
parameters are available in the literature, NGC\,6946 can be seen to
distinguish itself in several ways. The unusually strong FIR (Devereux
\& Young 1993) is likely to be associated with dust which is
coextensive with the molecular gas (Davies et al. 1999). The 850$\mu$m
image of Bianchi et al. (2000), shows, at least on the eastern side,
that the cold dust distribution is qualitatively very similar to that
of the mid-IR emitting dust and the molecular gas.

Although the total gas mass and total fractional gas mass, while high,
fall within the range seen in Sc galaxy samples (e.g. Schombert et al.
2001 and references therein), the molecular gas mass is an order of
magnitude higher than is typical (Boselli et al. 1997). This latter
result holds even when normalised by the dynamical mass (c.f. Fig. 2
of Casoli et al. 1998). The presence of so much molecular gas raises
questions such as where did it come from?  Why has it not formed
stars?

The idea that star formation is enhanced when the rate of cloud
collisions increases (Tan 2000) typically invokes an interaction as
the stimulus. Recently Karachentsev et al. (2000) and Pisano \&
Wilcots (2000) describe a small group of dwarf galaxies around
NGC\,6946 and Burton et al. (1999) reported the discovery of Cep 1, a
low-surface-brightness spiral galaxy in close spatial and kinematic
proximity to NGC\,6946. However, these studies find only low-mass
objects with no obvious signs of interaction with NGC\,6946. Placing
NGC\,6946 on the colour-asymmetry diagrams of Conselice et al. (2000)
argues against an interaction-driven trigger for the starburst in
NGC\,6946. An alternative to a single major interaction, is the
possibility of a series of minor mergers of gas-rich dwarf galaxies
(e.g. Bekki 1998; 2001: Rudnick et al. 2000). However it would require
flurry of some tens of typical dwarfs (Swaters 1999) to descend upon
NGC\,6946, even assuming that all of their neutral gas were to join
the molecular ISM.

A rather more speculative alternative arises out of the work of Walker
\& Wardle (1998; 1999) and Wardle \& Walker (1999), who revisit the
ideas of Pfenniger et al. (1994) and others, to imagine Galactic halo
dark matter consisting of cold, dense clouds of molecular hydrogen and
atomic helium. NGC\,6946's isolation may have allowed a relatively
fragile baryonic halo to have survived until some recent event, such
as a minor merger, heated and disrupted some fraction of the cold halo
clouds, which are known to be vulnerable to collisional disruption,
distributing them into the disk's ISM. This gas would settle into a
disk on an orbital time-scale. A problem with this scenario is that
the disrupted gas would initially be expected to be in atomic
form. Placing NGC\,6946 on the dark matter scaling relations of
Broeils (1992) (using the relatively well-determined parameter, the
dark halo mass within seven disk scale lengths (Walsh 1997)), one
finds that NGC\,6946's dark halo is significantly less massive than
expected for a galaxy of its maximum rotation speed.

Star formation efficiency (SFE) in a galaxy is generally defined as
the ratio of the mass of recently formed stars to the mass of
molecular gas from which they are presumed to have formed. The SFE may
be estimated by using H$\alpha$ and/or infrared luminosities to trace
the young stars and $^{12}$CO(1-0) observations to probe the molecular
gas mass. Both the mean star formation rate and the overall SFE for
NGC\,6946 are close to the mean value for Scd galaxies (Young et
al. 1996; Kennicutt 1998; Casoli et al. 1998). Rownd \& Young (1999)
using H$\alpha$ and $^{12}$CO(1-0) observations to consider the SFE in
a large sample of spiral galaxies, do not observe large SFE gradients
across star-forming disks, but their data have at at most a few
detected positions across any one disk.

\begin{figure}[htb]
\resizebox{\hsize}{!}{\includegraphics*[angle=0,totalheight=16cm]
{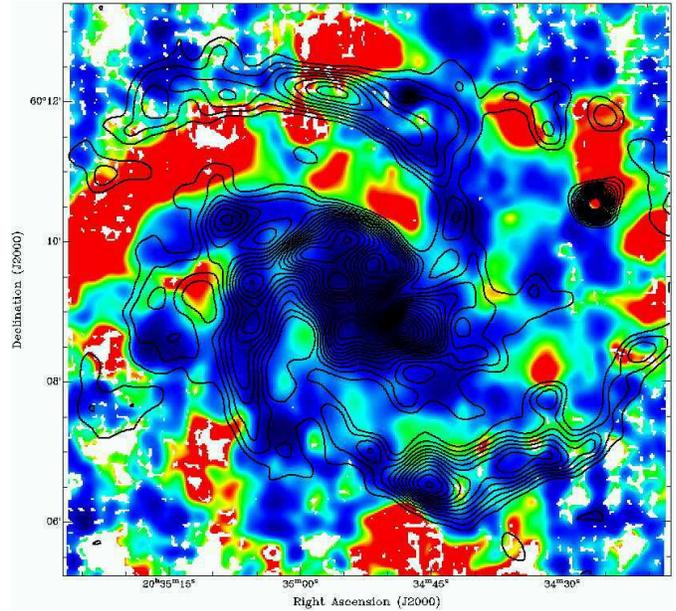}}
\caption[]{The false colour image shows the local star formation efficiency in NGC\,6946, formed from the ratio of the H$\alpha$ images to the $^{12}$CO(1-0) data, after both had been smoothed to the same spatial resolution. The contours, showing the 6cm polarised emission from 0.2 to 0.9 of the peak, in 5\% steps (since the edge of the polarised emission is very sharp, choice of contour levels is not critical), represent the regular part of the magnetic field in NGC\,6946.  \label{f:sfe}}
\end{figure}

We show in Fig.~\ref{f:sfe} the local SFE in NGC\,6946 with the
$\lambda$6\,cm polarised radio emission, representing the regular part
of the magnetic field, overlaid as contours. The SFE has been
estimated by dividing the H$\alpha$ image by the $^{12}$CO(1-0) image,
after selecting points of greater than $3\sigma$ significance and
smoothing to the same spatial resolution: this image is shown in
false colour in Fig.~\ref{f:sfe}. Fig.~\ref{f:sfe}, an important and
novel result, has several remarkable features. In Fig.~\ref{f:sfe} the
SFE ranges by over two orders of magnitude with highest values found
in the northeastern spiral arm. Low values are found in the central
regions, but it is not clear to what extent this is due to the high CO
brightnesses combined with the high central optical extinction
(c.f. Fig. 2 of Devereux \& Young 1993). Beyond the high extinction
region, considerable variations of SFE occur, with high values
coincident with the outer spiral arms. That the SFE is generally
greater at larger radii is not unprecedented in the sample of Rownd \&
Young (1999), but the maximum variation in SFE is larger than in any
of their sample.

A striking feature of Fig.~\ref{f:sfe} is the way in which the
contours representing the regular part of the magnetic field trace out
regions of low SFE. Polarised emission is strong both north and south
of the northeastern spiral arm, but entirely absent within it. The
polarised emission even traces out two regions of low SFE within the
arm (at $\alpha = 20^{\rm h}\, 34^{\rm m}\, 59.5^{\rm s}, \delta =
60^{\circ} 11\arcmin\, 16\arcsec\, $ and $\alpha = 20^{\rm h}\,
34^{\rm m}\, 42.0^{\rm s}, \delta = 60^{\circ} 10\arcmin\, 12\arcsec\,
$ (J2000)). Regions of higher SFE also avoid the polarised emission in
the south. We performed a two-dimensional correlation analysis on the
SFE and $\lambda$6\,cm polarised emission images using the methodology
described in Sect.~\ref{s:twodim}.  The result of a linear correlation
analysis is a regression line with slope $-0.16\pm0.6$ and a Pearson's
correlation coefficient of $-0.3\pm0.1$. The analysis is sensitive to
a few high points in the SFE image, and when these points are excised,
the slope decreases by a factor of three, the correlation coefficient
becomes $-0.4\pm0.1$. However in both cases the significance of the
correlation is low, with the Student's $t$ value less than 4. The
significance of the correlation is only slightly increased ($t \sim
5$) when the correlation is performed logarithmically. Thus the
apparent strong anticorrelation is possibly coincidental, given that
the definition of the ``regular'' part of the magnetic field is
related to the resolving beamsize and is not an intrinsic property of
the galaxy. The appearance of Fig.~\ref{f:sfe} may be due to SF
tangling the field within the spiral arms, the dynamo producing
regular fields preferably in interarm regions where turbulence and/or
diffusivity is smaller (c.f. Rohde et al. 1999), or a hitherto
unsuspected relationship between large scale magnetic fields and
molecular cloud and/or star formation.

\section{Conclusions}
\label{s:conclusion}

\begin{enumerate}
\item We have mapped the $^{12}$CO(1-0) and $^{12}$CO(3-2) emission in NGC\,6946, using the 
IRAM 30-m radio telescope and the Heinrich Hertz Telescope, respectively. The beam width of 22\arcsec\, at 115\,GHz and 345\,GHz corresponds to a linear size of
590\,pc at the assumed distance of 5.5\,Mpc. Extended CO emission is imaged in both lines across the entire disk of this gas--rich, actively star--forming galaxy.
\item We find that warm, dense gas appears to be distributed over an area of some 3\,kpc in NGC\,6946 and is also located at non-central positions. The $^{12}$CO(3-2) images contain considerable diffuse emission between hotspots and spiral arms indicating that molecular clouds containing warm and dense gas are distributed throughout the inner disk of NGC\,6946.
\item The arm--interarm contrast in two selected areas of the NE spiral arm 
is $1.2 \pm 0.2$ in $^{12}$CO(1-0), similar to that of the {H\,{\sc i}}\ gas,
and $1.8 \pm 0.2$ in $^{12}$CO(3-2), measured over the same areas. The larger contrast of $^{12}$CO(3-2) suggests that molecules in the spiral
arms are warmer or reside for longer in the spiral arm.
\item The rotation curve derived from the $^{12}$CO(1-0) velocity field agrees,
within the uncertainties, with the {H\,{\sc i}}\ rotation curve of Carignan
et al. (1990). The velocity dispersion is $40\pm10$\,km\,s$^{-1}$ in the inner 2\,kpc and $8\pm3$\,km\,s$^{-1}$ in the disk, without tendency to increase in the spiral arms.
\item We find that the total molecular gas mass is exceptionally massive, 
with M$_{\mathrm{H}_2} = 1.13 \times 10^{10}\, \mathrm{M}_{\odot}$, which is 
about an order of magnitude more than that of both the Galaxy and the median for the large sample of spiral galaxies studied by Casoli et al. (1998), and thus confirm that the molecular component in NGC\,6946 is almost as massive as the atomic gas, with $\mathrm{M}_{\mathrm{H}_2}/\mathrm{M}_{\mathrm{HI}} = 0.57$. 
\item We have observed the excitation conditions as a function of location
of the gas clouds relative to the star-forming regions in
NGC\,6946. Using a large-velocity gradient code, we modeled the
temperature and density of the gas and find that, in the centre of
NGC\,6946 the beam-averaged gas kinetic temperature is $40\pm5$\,K and
the molecular gas density is $(3.3\pm0.3) \times 10^3$\,cm$^{-3}$. The
optical depth of the $^{12}$CO(3-2)  line is high, with $\tau_{^{12}\mathrm{CO}(3-2)}\sim 10$.
\item We find a lack of association between the large-scale
distributions of the main molecular gas tracer, $^{12}$CO(1-0), and the atomic
gas. Taken together with the considerable star formation in the disk, this
argues against the hypothesis that {H\,{\sc i}}\ is the product of 
H$_2$ photodissociation by UV radiation from young stars.
\item The highest correlation between any pair of tracers is found between
the 7\,$\mu$m MIR emission and the total radio continuum emission
at $\lambda$6\,cm. This cannot be due to dust heating and gas ionisation
in star-forming regions because the thermal radio emission is {\it less} 
correlated with the MIR emission than the nonthermal emission.
A coupling of magnetic fields to gas clouds is a possible scenario.
\item When the $\lambda$6\,cm polarised emission, which traces the regular part of the magnetic field, is overlaid on an image made from the smoothed H$\alpha$ map divided by the molecular gas map (the SFE image), it is seen that the magnetic field traces out regions of low SFE.
\end{enumerate}

\begin{acknowledgements}

We thank the staff at IRAM and SMTO for extensive help with the
observations. We are grateful to S. Larsen, D. Dale, A. Ferguson,
R. Dettmar, M. Trewhella and R. Kamphuis for providing us with their
datasets. It is a pleasure to thank E. Berkhuijsen, C. Henkel,
A. Schulz, A. Vallenari, M. Walker and an anonymous referee for
discussions that added to the content of this paper, and C. Nieten for
help with his correlation software.

\end{acknowledgements}

\end{document}